\vskip 5pt
\vskip 5pt
\centerline{\bf Unified theory of elementary fermions and their interactions based on Clifford algebras} 
\vskip 5pt

\centerline{Douglas Newman}
\vskip 5pt
\centerline {e-mail: \it dougnewman276@gmail.com}
\vskip 10pt

\beginsection Abstract

Seven commuting elements of the Clifford algebra $Cl_{7,7}$ define seven
binary eigenvalues that distinguish the $2^7=128$ states of 32 
fermions, and determine their parity, electric charge and interactions. 
Three commuting elements of the sub-algebra $Cl_{3,3}$ define three binary quantum numbers
that distinguish the eight states of lepton doublets.   
The Dirac equation is reformulated in terms of a Lorentz invariant operator 
which expresses the properties of these states in terms of Dirac 4-component
spinors. Re-formulation of the Standard Model shows chiral symmetry breaking to be redundant. 
A $Cl_{3,3}$ sub-algebra of $Cl_{5,5}$ defines two additional binary quantum numbers 
that distinguish quarks and leptons, and describes the SU(3) gluons 
that produce the hadron substrate, explaining quark confinement.  
Finally, a $Cl_{3,3}$ sub-algebra of $Cl_{7,7}$ defines a further 
two binary quantum numbers that distinguish four fermion generations. The predicted fourth  
generation is shown to have no neutrino and a distinct substrate, suggesting that 
ordinary matter is confined and providing candidates for unconfined dark matter. 
Interactions between fermions in the first three generations are predicted, including
those that produce flavour symmetry. Relationships are explored between the $Cl_{1,3}$ algebra 
and general relativity, and between $Cl_{5,5}$ and SO(32) string theory.

\vfill\eject

\beginsection \S1: Introduction 

The  main features of the Standard Model (SM) were formulated between 
1961 and 1967 (e.g. see Appendix 6 of [1]), producing a comprehensive       
conceptual and mathematical model of elementary
particles and their interactions that is generally accepted as 
providing excellent agreement with experiment. 
Nevertheless, it lacks a coherent formalism, which limits its 
predictive capability and (as will be shown in this work) invalidates 
some of this `agreement'.

From 1974 onwards, many attempts were made to extend the SM 
formalism by employing Lie groups which have, as sub-groups, the 
SU(2) and SU(3) gauge groups that describe weak and strong interactions. 
Particular attention, summarized in [2,3], was given to SU(5)   
and SO(10). A great deal of effort, often centred on super-symmetry concepts [4], 
has since been expended in trying to repair the defects 
in these early attempts at unification. In retrospect, their problems
arose because they incorporated the mathematical formalism of the
SM, including the role of chirality, in their description of the 
elementary fermions. Clifford Unification is based on a new algebraic description
of all the elementary fermions, which replaces the SM. The unification it 
achieves should not be confused with past attempts to unify gauge fields.

String theory [5] and Clifford algebras share a common interest           
in higher dimensional metrics. Their study originated with 
the Kaluza-Klein unification of gravity and electro-magnetism 
by extending the space-time metric to five-dimensions. String theory is 
based on the discovery that a ten-dimensional 
space-time metric had attractive mathematical
properties that could be used to describe elementary bosons and fermions. 
In spite of the tremendous effort that has been devoted to the elaboration of 
its formalism, no clear relationship between 
the theoretical constructs of string theory
and particle physics has been found. 

Eddington [6] realized that the Dirac algebra could be employed              
as a  common basis for the description of classical mechanics, gravitation
and relativistic quantum physics. Unfortunately, there was little relevant experimental
data at that time, and his personal attempt to predict elementary
particle properties has made this approach a no-go area for generations of physicists.
Nevertheless, the value of $Cl_{1,3}$ algebra in the description of space-time is 
now well established, e.g [7,8]. It has been known since 1958 that this     
algebra puts Maxwell's equations in vacuo into a particularly simple form [8,9]
related to the Dirac equation for zero mass fermions, but it has not
been possible to find a Clifford algebra that provides a coherent link between 
space-time algebra and the description of fermions. 

In 2001 Trayling and Baylis [10] identified the SU(2) and SU(3)  
Lie algebras in $Cl_7$. In 2009, Dartora and Cabrera [11] showed that the main 
features of electro-weak theory can be explained in terms of the $Cl_{3,3}$ 
algebra if chirality is omitted. The present work incorporates several 
of their results. Unfortunately,
their misidentification of the time coordinate, and (possibly) the 
characterization of their work as a 'toy' theory in the abstract, 
has led to their work being ignored. The interpretation of elementary particle 
properties in terms of $Cl_{6,0}$ as a description of non-relativistic phase space by 
Zenczykowski [12,13,14], is also relevant. More recently, Stoica [15,16] 
has shown that the results in [10] can also be expressed
using the complex Clifford algebra $Cl^*_6$, and has investigated 
how this algebra might incorporate chiral symmetry breaking.  
It would be of interest to relate these approaches
to the $Cl_{n,n}$ algebras, but this has not been 
attempted in this work.                                                          

Pav\v si\v c [17] has given string theoretic arguments for                             
the importance of $Cl_{8,8}$ in providing a description of 
the elementary fermions. Yamatsu [18] has described a grand unified theory based on the     
Lie group USp(32), which is related to SO(32) string theory. 
Given that the Lie algebra of SO(32) and $Cl_{5,5}$ are both algebras
of $32\times 32$ matrices, there are possible links between Yamatsu's work and 
the present work. 

Although the present theory does not incorporate the 
algebraic structure of the SM, some detailed
comparisons have been necessary. These have been helped by the 
many excellent textbooks on the SM that are now available. 
These include the thorough theoretical approach in 
Aitchison and Hey [19,20] and the clarity of presentation provided by Thomson [21].
The recent edition of the book by Dodd and Gripalos [22] has also been useful.

\vfill\eject

\beginsection \S2. Procedure

Clifford algebras were originally developed in the context of 
algebraic geometry, and are particularly appropriate for 
the description of macroscopic observables
in a way that is independent of the observer's coordinate system [7,8].           
The main reason for thinking that they
could provide useful models of elementary 
fermions and their interactions is the
role played by $Cl_{1,3}$ in the Dirac equation, where 4-spinors 
both distinguish fermion states and describe their dynamics. 
The successful application of the Dirac equation in quantum electrodynamics
makes it clear that its algebra must provide the core 
of any unified theory. Hence the algebras 
studied in this work necessarily contain $Cl_{1,3}$ as a sub-algebra.
The choice of algebras is dependent on maintaining precise relations
between their algebraic structures and the interpretation
of observations. This work is concerned with identifying the discrete
properties that distinguish elementary fermions and bosons, while 
keeping the successful aspects of the Dirac equation and Standard Model intact.  
Unification is developed in three stages, corresponding to 
the Clifford algebras $Cl_{3,3}\subset Cl_{5,5} \subset Cl_{7,7}$.
The quantum numbers obtained at each stage are given physical interpretations
in terms of the elementary fermions and their interactions with
gauge fields, as follows:\vskip 5pt

{\bf Stage 1: Lepton properties based on $Cl_{3,3}(L)$}

\item {\S3,1} Summarises the geometrical interpretation of the
$Cl_{1,3}$ space-time algebra. 

\item {\S3,2} Introduces a real 8$\times$8 matrix 
representation of $Cl_{1,3}$ and extends this
to a representation of $Cl_{3,3}$. {\it Time intervals are 
identified as the product of all six generators of $Cl_{3,3}$.}   
 
\item {\S3,3} Interprets the algebraic expression for Maxwell's field
equations in vacuo as a photon wave-equation, with
wave-functions expressed as excitations of a specific substrate.

\item{\S4,1} Describes eight lepton states in
terms of three commuting elements of $Cl_{3,3}$, with eigenvalues
corresponding to binary quantum numbers that provide a formula 
for lepton charges.

\item {\S4,2}  Relates the physical properties of leptons 
to the seven Lorentz invariants defined by the commuting 
elements of $Cl_{3,3}$.

\item {\S4,3} Derives the effect of discrete
 coordinate transformations on lepton properties.

\item {\S5,1} Reformulates the Dirac equation as a Lorentz
 invariant differential operator acting on a Lorentz invariant, 
 avoiding the negative mass problem.

\item {\S5,2} Reformulates the SM description of the Higgs boson
while keeping its physical interpretation. 

\item {\S5,3} Relates the differential operator to canonical
momentum, showing that fermion properties are determined by the substrate 
of their wave motion, rather than their internal structure. 

\item{\S6,1} Expresses the weak interaction in terms 
of the generators of $Cl_{3,3}$, formulating 
electron/neutrino interactions {\it without reference to chirality}.

\item {\S6,2} Shows the $Cl_{3,3}$ formulation of the weak interaction
gives opposite parities of electron and neutrino spatial coordinates .

\item {\S6,3} Revises the Standard Model integration of electromagnetic
and weak interactions. 

\vskip5pt {\bf Stage 2: Quark and lepton properties based on $Cl_{5,5}(LQ)$}

\item {\S7,1} Relates $Cl_{5,5}$ generators to those of $Cl_{3,3}(L)$,
determining two additional quantum numbers extending the formula
for fermion charges to include quarks.

\item {\S7,2} Defines $Cl_{3,3}(Q)$, showing the SU(3) 
Lie algebra to be a sub-algebra of $Cl_{5,5}(LQ)$.

\item {\S7,3} Interprets quark properties in terms of a gluon jelly substrate. 

\vskip5pt{\bf Stage 3: $Cl_{7,7}$}

\item {\S8,1} Relates $Cl_{7,7}$ generators to those of $Cl_{3,3}(L)$ and $Cl_{5,5}(LQ)$,
determining two additional quantum numbers, giving seven overall, extending 
the formula for fermion charges to include four generations, and showing the fourth
generation to have no neutrino.

\item {\S8,2} Distinguishes the substrate of the fourth predicted 
generation from that of the three known generations.

\item {\S8,3} Identifies possible gauge fields and elementary bosons 
that are consistent with the algebra.

\item {\S8,4} Discuses the observability of the predicted fourth generation
of fermions. 

\vskip 8pt 

\S9 outlines the relationship between the formalism 
and general relativity. \S10 identifies a relationship 
with string theory. \S11 discusses the {\it substrate} concept.

\beginsection \S3. From space-time algebra to $Cl_{3,3}$

The Clifford space-time algebra $Cl_{1,3}$ has four anti-commuting
generators, denoted ${\bf E_\mu},\>\{\mu=0,1,2,3\}$, interpreted as
unit displacements in the four coordinate directions. They satisfy
$$
{\bf E_\mu}{\bf E_\nu}+{\bf E_\nu}{\bf E_\mu} = 2 g_{\mu\nu}, \eqno(3.1)
$$
where the Minkowski metric tensor $g_{\mu \nu}$ has zero
components when $\mu \not= \nu$ and 
$$
g_{11}=g_{22}=g_{33}= -1,\> g_{00}=1,\>{\rm so \>that}\> \>
g_{\mu\mu}= ({\bf E_\mu})^2.                                  \eqno(3.2)
$$
Raising and lowering suffices follows the tensor convention, i.e.
${\bf E^\nu}= g^{\nu\mu}{\bf E_\mu}$.
Combining the ${\bf E_\mu}$ with rank 1 tensors produces Lorentz invariant
expressions called {\bf structors} in this work. These are to be 
distinguished from those single elements of the $Cl_{3,3}$ algebras
that are themselves Lorentz invariant.
For example, infinitesimal displacements in space-time are expressed as
the structor 
$$
d{\bf x}={\bf  E}_\mu dx^\mu,                                 \eqno (3.3)
$$
where it is assumed that all four unit displacements 
have the same dimensions (e.g. centimetres).  
$d{\bf x}^2>0$ for displacements of particle with 
finite mass and $d{\bf x}^2=0$ for photons.

Orientated unit areas in space-time are expressed as
$$
{\bf E}_{\mu \nu}= {1\over 2}({\bf E}_\mu {\bf E}_\nu - {\bf E}_\nu
{\bf E}_\mu),                                                          \eqno (3.4)
$$
so that infinitesimal area structors have the form
$$
d^{\,2} {\bf S}= {\bf E}_{\mu\nu}dx^\mu dx^\nu.                         \eqno (3.5)
$$ 
Similarly, unit 4-dimensional volumes are defined in terms of the
element denoted ${\bf E}^\pi$ of the $Cl_{1,3}$ algebra, i.e.
$$
{\bf E}^\pi = {\bf E}_0 {\bf E}_1 {\bf
	E}_2 {\bf E}_3 = {1\over 4!} \epsilon^{\mu \nu \kappa \tau}{\bf
	E}_\mu {\bf E}_\nu {\bf E}_\kappa {\bf E}_\tau .                     \eqno (3.6)
$$
(The suffix $\pi$ does not take numerical values.) The anti-symmetizer
$\epsilon^{\mu \nu \kappa \tau}$ is zero if any two suffices are equal, $+1$
for suffices that are even permutations of $\{0,1,2,3\}$, and $-1$
for suffices that are odd permutations of $\{0,1,2,3\}$. 
Infinitesimal space-time volumes ${\bf v}_4$ therefore correspond to the structor
$$
d^{\,4}{\bf v}_4= {\bf E}^\pi\, d\tau = {1\over 4!}\,{\bf E_\mu}{\bf
	E_\nu}{\bf E_\kappa}{\bf E_\rho} \, dx^\mu dx^\nu dx^\kappa dx^\rho.  \eqno (3.7)
$$
Three-dimensional unit `surface areas' are given by the 
products 
$$
{\bf E}^{\pi\tau} = {\bf E}^\pi {\bf E}^\tau = {1\over 3!}\epsilon^{\mu
	\nu \kappa \tau}{\bf E}_\mu {\bf E}_\nu {\bf E}_\kappa .                \eqno (3.8)
$$
In particular, ${\bf E}^{\pi0}$ is the unit spatial volume.
Infinitesimal 3-dimensional volumes have the structor form
$$
d^{\,3}{\bf S}= {\bf E}^{\pi\tau} dS_\tau ={1\over 3!}{\bf E_\mu}{\bf
	E_\nu}{\bf E_\kappa} dx^\mu dx^\nu dx^\kappa.                           \eqno (3.9)
$$
The number of elements in a Clifford algebra determines how
many different physical constructs can be described in terms of
measurements of the unit displacements defined by its generators.
A consequence of this is that when physical laws are expressed in terms of
structors, the {\it closure} of $Cl_{1,3}$ constrains their form 
in a way that goes beyond Lorentz covariance. An important example is
$$	
{\bf E}_{\mu\nu} {\bf E}_\kappa =\epsilon_{\mu \nu \kappa\tau}
{\bf E}^{\pi\tau} + g_{\nu\kappa}{\bf E}_\mu - g_{\mu\kappa}{\bf E}_\nu.      \eqno(3.10)
$$

The  Lorentz invariant differential operator is the structor 
$$
{\bf D} = {\bf E^\mu} {\partial}_\mu.                                        \eqno (3.11)
$$
Its geometrical interpretation is provided by  the integral operator equality
$$
\int_{\bf v} {d^4{\bf v}\>{\bf D}}{\bf X}=
\int_{S{\bf v}}{ d^3{\bf S}}\> {\bf X},                                        \eqno (3.12)
$$
where the 4-volume and 3-surface structors are given above. This 
is a special case of the Boundary Theorem (e.g. [7], p.69).                  
The structor ${\bf X}$ in (3.12) is arbitrary,
the integral on the left hand side is taken over a 4-volume $\tau$,
and the integral on the right hand side is taken over the
3-dimensional surface $S(\tau)$ that encloses the 4-volume.  

Transformations $\bf\Lambda$ relating structural coefficients in
different Minkowski reference frames, denoted ${\bf E}^\nu$ and ${\bf F}^\mu$, 
can be expressed either as a similarity transformation or as 
a linear relationship between the coordinates, viz.
$$
{\bf F}^\mu ={\bf\Lambda}{\bf E}^\mu {\bf\Lambda}^{-1} = {\bf E}^\nu  \Lambda_{\nu}^\mu.  \eqno (3.13)
$$
The $\Lambda_{\nu}^\mu$ express the transformation in terms of rotations
of the spatial coordinates ${\bf E}_1,\>{\bf E}_2,\>{\bf E}_3$, and boosts relating
the spatial coordinates to ${\bf E}_0$. Its algebraic form 
has been analysed in great detail, e.g. in [8], but is not relevant to this work.

Structors are also subject to  discrete transformations that
cannot be expressed as Lorentz transformations. 
As these are often involved in the analysis 
of elementary particle interactions it is necessary to establish their 
algebraic form. The spatial inversion, or parity, transformation  $\bf \hat P$
changes the sign of all three spatial coordinates in a specific reference frame, 
and the sign of the unit spatial volume  ${\bf E}^{\pi0}$, i.e.
$$
{\bf E^\mu}\rightarrow {\bf \hat P} {\bf E^\mu}{\bf \hat P}^{-1}= {\bf E_\mu}, 
\>\>{\rm where}\>\>{\bf \hat P} = {\bf \hat P}^{-1} = {\bf E^0}.                           \eqno (3.14)
$$
This transformation, and reflections, which 
change the sign of any one of 
$ {\bf E}_1,\> {\bf E}_2,\> {\bf E}_3$, interchange right and left
handed spatial coordinate systems, so that
${\bf E}^{\pi0} = {\bf E}_1 {\bf E}_2 {\bf E}_3\rightarrow -{\bf E}^{\pi0}$ 
and ${\bf E}^{\pi} = {\bf E}^{\pi0} {\bf E}^{0}\rightarrow -{\bf E}^{\pi}$.  
Coordinate time inversion $ {\bf \hat T} = {\bf E}^{\pi 0}$
changes the sign of ${\bf E}^{0}$, corresponding to running clocks backwards,
 without changing the spatial coordinate directions, so that
$$
{\bf E^\mu}\rightarrow {\bf \hat T}{\bf E^\mu}{\bf \hat T}^{-1} = -{\bf E_\mu}.                 \eqno (3.15)
$$
Proper time inversion $ {\cal T}={\bf \hat T \hat P}= \bf {\hat P\hat T}=\bf E^{\pi}$,  
changes the sign of all the ${\bf E^\mu}$ in any reference frame, giving
$$
{\bf E^\mu}\rightarrow {\cal T} {\bf E^\mu}{\cal T}^{-1} = -{\bf E^\mu},                     \eqno (3.16)
$$

While particles have instantaneous positions in space,
relativity theory expresses them as structors describing their 
infinitesimal displacements (3.3) in space-time. These take
a special form in the rest frame of massive particles, i.e.
$$\eqalign{
d{\bf x}=\>{\bf E}_{*0} dx^{*0}= &\> {\bf  E}_\mu dx^\mu,\> \mu = 0,1,2,3
\>\>{\rm so \>\>that } \>\> {\bf E}_{*0}= {\bf  E}_\mu {{dx^\mu}\over dx^{*0}}\cr 
{\rm giving}\>\>(d{\bf x})^2 =&\> ({\bf E}_{*0} dx^{*0})^2 = (dx^{*0})^2,
{\bf D}={\bf E}^{\mu}\partial_{\mu} = {\bf E}^{*0} \partial_{*0},\>\>{\rm and \>}\cr                                                     
{\rm the\>momentum}  \> {\bf p}=&\> m {\bf E}_{*0} = m{\bf  E}_\mu 
{{dx^\mu}\over dx^{*0}},\>\>{\rm where} \> m {\rm \>is\> the\>particle \> mass}.}                                                                               \eqno (3.17)
$$
Here the `star' in  ${\bf E}_{*0}={\bf E}^{*0}$ and $dx^{*0}$ distinguishes 
between proper time intervals, measured in the rest frame of the particle, 
from time intervals ${\bf E_{0}}dx^0$ in an arbitrary reference frame. 
In relativistic classical mechanics the magnitude $dx^{*0}$ of a 
particle's displacement in space-time is often written $ds$. 
The `star' notation will also be used to distinguish between spatial 
displacements in the particle and observer's reference frames. 
It will only be necessary to make this distinction, 
i.e. introducing all the particle frame components ${\bf E^{*\mu}}$, 
when physical descriptions relate to arbitrary reference frames. 
The main role of the particle frame is that its geometry, 
i.e. $\pm$ spin and the time direction,
form part of the invariant description of fermions.   
All structors have scalar magnitudes determined by their square, 
which can be positive, negative or zero.
This will sometimes be made explicit by putting $(\pm)$ or (0) after the label.  

In classical mechanics particles are conceived as the stable and single 
occupants of points in 3-dimensional space. Their dynamical properties 
are mass, electric charge, velocity and kinetic energy. 
$Cl_{1,3}$ space-time geometry, as outlined above, provides all 
that is necessary to describe their dynamics, making it unnecessary to introduce
matrix representations (as pointed out in [8]). However, matrix representations 
are necessary for the description of fermions.

The first step in relating the Dirac-Pauli matrix representation of $Cl_{1,3}$
to the interpretation of the same algebra in classical mechanics is to obtain 
a real $\gamma$-matrix representation. In order to distinguish the two representations  
the notation $\bar\gamma$ is used for the Dirac-Pauli matrices. 
Given that the required $\gamma$-matrix representation is real, 
and to distinguish algebraic and scalar  
occurrences of the square roots of $-1$ in the following analysis,
both sets of matrices will be expressed 
in terms of the four linearly independent {\it real} 2$\times$2 matrices, 
$$
{\bf I} =
\left(\matrix{1&0\cr
	0&1\cr}\right),\>
{\bf P} = -i\sigma_2 = \left(\matrix{0&-1\cr
	1&0\cr}\right),\>
{\bf Q} = \sigma_1 =\left(\matrix{0&1\cr
	1&0\cr}\right),\>
{\bf R} = -\sigma_3 =\left(\matrix{-1&0\cr
	0&1\cr}\right),                                                     \eqno (3.18)                        
$$
where the $\sigma$s denote Pauli matrices. The 2$\times$2 matrices satisfy 
$$
{\bf P}{\bf Q}={\bf R},\>\>-{\bf P}^2={\bf Q}^2 = {\bf R}^2 = {\bf I}.    \eqno (3.19)
$$

The  generators of the $Cl_{1,3}$ Dirac algebra can be expressed as Kronecker products, viz.
$$
\bar\gamma^0 = -{\bf I}\otimes{\bf R},\>\> \bar\gamma^1= -{\bf Q}\otimes{\bf P},
\> \>  \bar\gamma^2 = -i{\bf P}\otimes{\bf P}, \>\> \bar\gamma^3={\bf R}\otimes{\bf P},\eqno(3.20)
$$
and an additional matrix is defined as 
$$
\bar{\gamma}^5 = i\bar\gamma^0 \bar\gamma^1 \bar\gamma^2 \bar\gamma^3 = {\bf I}\otimes{\bf Q}. \eqno (3.21)
$$  
No real 4$\times$4 matrix representation of the $Cl_{1,3}$ algebra exists, but a real 8$\times$8
representation can be constructed. Its generators, defined in (3.20),
are mapped into their corresponding real representation matrices
by adding a factor $\otimes {\bf I}$ to the real
$\bar\gamma$ matrices, and replacing the factor 
$i$ in $\bar\gamma^2$ by $-{\bf I}\otimes{\bf I}\otimes{\bf P}$. In summary
$$\eqalign{ 
{\bf I}\otimes{\bf I}&  \rightarrow {\bf 1}_3 ={\bf I}\otimes{\bf I}\otimes{\bf I},
\>i{\bf I}\otimes{\bf I}\rightarrow \gamma^{\pi 6} =-{\bf I}\otimes{\bf I}\otimes{\bf P},\cr
        \bar\gamma^0 &  \rightarrow \gamma^0 = -{\bf I}\otimes{\bf R}\otimes{\bf I},
      \>\bar\gamma^1    \rightarrow \gamma^1= -{\bf Q}\otimes{\bf P}\otimes{\bf I},\cr
      \>\bar\gamma^2 &  \rightarrow  \gamma^2 = {\bf P}\otimes{\bf P}\otimes{\bf P}, 
      \>\bar\gamma^3    \rightarrow  \gamma^3={\bf R}\otimes{\bf P}\otimes{\bf I},}     \eqno (3.22)
$$
where $\gamma^\mu$ is the real matrix representation of ${\bf E}^\mu$.  Space-time unit volumes 
are 
$$
\gamma^\pi = \gamma^0 \gamma^1 \gamma^2 \gamma^3 = {\bf I}\otimes{\bf Q}\otimes{\bf P}, \eqno(3.23)
$$ 
which does {\bf  not} correspond to $\bar\gamma^5$ in (3.21). The 8$\times$8
 matrix corresponding to $\bar\gamma^5$ is  
$\gamma^6 = {\bf I}\otimes{\bf Q}\otimes{\bf I}$, 
identified in Table A1 as one of the three time-like generators of $Cl_{3,3}$. 

Products of the $\gamma^\mu\> \{\mu= 0,1,2,3,6\}$ generate the 32 entries 
in the second and third columns of Table A1 of 
Appendix A. The complete table has 64 matrices, providing 
a real representation of the $Cl_{3,3}$ algebra. It is
obtained by introducing the time-like generators 
$\gamma^7= -{\bf P}\otimes{\bf P}\otimes{\bf Q}$ 
and $\gamma^8={\bf P}\otimes{\bf P}\otimes{\bf R}$, 
which anti-commute with all four generators of $Cl_{1,3}$. 
The six matrices $\gamma^\mu,\>\{\mu=1,2,3,6,7,8\}$
provide all six generators of $Cl_{3,3}$, 
with unit space-time displacements denoted 
$\gamma^{\mu},\>{\rm where}\>\{\mu= 1,2,3,0\}$. 
$\gamma^\pi$ and the generators $\gamma^{\mu},\>{\rm where}\>{\mu= 6,7,8}$, 
are Lorentz invariant. Unit time displacements do not appear as one of the 
generators of $Cl_{3,3}$ but
are given by $\gamma^0 =\gamma^1\gamma^2\gamma^3\gamma^6\gamma^7\gamma^8$.
This can be simplified by noting that $\gamma^6\gamma^7\gamma^8 = \gamma^\pi$,
showing that units of time correspond to the 
unit space-time 4-volume divided by its corresponding unit spatial 3-volume. 

In the remainder of this paper it will be assumed that all elements 
${\bf E}^\mu$ of the Clifford algebras have matrix representations, 
and the same notation will be used for structors
and their matrix representations. The canonical matrix representation 
of the electromagnetic field structor in vacuo is  
$$\eqalign{
 {\bf F} = &\gamma^{\mu\nu} F_{\mu\nu}/2 \cr
 = &\left(\matrix{
	0&-F_{31}&F_{03}&F_{01}&-F_{12}&-F_{23} &0&-F_{02}\cr
	F_{31}&0&F_{01}&-F_{03}&-F_{23} & F_{12}&F_{02}&0\cr
	F_{03}&F_{01}&0&-F_{31}&0&-F_{02}&-F_{12}&-F_{23}\cr
	F_{01}&-F_{03}&F_{31}&0& F_{02} &0&-F_{23}&F_{12}\cr
	F_{12}&F_{23}&0&F_{02}&0&-F_{31}&F_{03}&F_{01}\cr
	F_{23}&-F_{12}&-F_{02}&0&F_{31}&0&F_{01}&-F_{03}\cr
	0&F_{02}&F_{12} & F_{23}&F_{03}& F_{01}&0 & -F_{31}\cr
	-F_{02} &0& F_{23}& -F_{12}&F_{01}&-F_{03}&F_{31}&0\cr}\right) \cr
&\cr
 = &[ {\bf F}_{(1)},\>{\bf F}_{(2)}= \gamma^{\pi7} {\bf F}_{(1)},\>
 {\bf F}_{(3)}= \gamma^6 {\bf F}_{(1)},\>{\bf F}_{(4)}= -\gamma^8 {\bf F}_{(1)},\cr
   &\>\>\>\>\>{\bf F}_{(5)}= \gamma^{\pi6}  {\bf F}_{(1)},
   {\bf F}_{(6)}= -\gamma^{\pi 8}{\bf F}_{(1)},{\bf F}_{(7)}= \gamma^\pi {\bf F}_{(1)}, 
   {\bf F}_{(8)}=-\gamma^7{\bf F}_{(1)}]}   ,                                            \eqno(3.24)
$$
where  ${\bf F}_{(i)}$ is the $i$-th column of ${\bf F}$. Maxwell's equations 
can be expressed by the structor equation
$$
 {\bf DF}= {\bf J},           \eqno(3.25)
$$
where the charge-current density structor ${\bf J} = J_\mu \gamma^\mu $ 
is the source of $\bf F$.  {\it (3.25) shows Maxwell's equations 
in vacuo to be a consequence the closure relation (3.10)}. 
In vacuo, each column of $\bf F$ separately 
satisfies ${\bf D}{\bf F}_{(i)}= 0$, as will column matrices 
formed from any linear combination 
${\bf\Phi}_F = \sum_i a_i {\bf F}_{(i)}$, where the coefficients 
$a_i$ are constant complex numbers. The equation
$$
{\bf D}{\bf\Phi}_F = 0     \eqno (3.26) 
$$
has the same structure as the Dirac wave-equation for particles 
of zero mass (after making the modifications described in \S4).
When $\bf F$ describes a radiative field, constraints on the magnitudes of the
electric and magnetic components of the field correspond to the structor 
equation ${\bf F}^2 = 0$, so the eight terms in the product of any row with any 
column of $\bf F$ sum to zero.  Given this constraint, and the adjoint 
$({\bf\Phi}_F)^\dagger $ of ${\bf\Phi}_F$, $({\bf\Phi}_F)^\dagger {\bf\Phi}_F = 0$,
{\it so that (3.26) provides a wave-mechanical description of photons}. 

Interactions between photons and fermions are conventionally formulated in terms of  potential  
structors $\bf A= A_{\mu} \gamma^{\mu}$, related to the electromagnetic field by
$$
{\bf F = DA} = \gamma^\mu \partial_\mu \gamma^\nu A_\nu 
= \gamma^{\mu\nu}(\partial_\mu A_\nu - \partial_\nu A_\mu)/2 + \partial_\mu A^\mu.                                                               \eqno(3.27)                                  
$$
It follows from that ${\bf F}_{(i)} = {\bf D}{\bf A}_{(i)}$, giving
$$
{\bf\Phi}_F = {\bf D}{\bf \Phi}_{\bf A}= {\bf D}\sum_i a_i {\bf A}_{(i)}  , \eqno(3.28)
$$
where the ${\bf A}_{(i)} $ denote columns of ${\bf A}$.
The conventional plane-wave description of photons has the structor form
$$
{\bf A}=  \exp ( \eta k_\mu x^\mu) {\bf A}^{\rm const.},                             \eqno (3.29) 
$$ 
where ${\bf k}=\gamma^\mu k_\mu,\>{\bf A}^{\rm const.}$ are independent of
the space and time coordinates $x^\mu$, and $\eta=i$. The identification 
$\eta = i$, accords with the Michelson-Morley result that no substrate 
for photon waves in the form of a stationary `aether' exists. This does 
not, however, rule out the possibility that photon wave motion 
modulates a medium that can be expressed algebraically in terms of
a Lorentz invariant $\eta$, providing a physical substrate in which 
the photons propagate. The following analysis is made on the basis 
that possible choices $\eta \not= i$, with $\eta^2 = -1$, exist.

It follows from (3.29) that
$$
{\bf F} = {\bf D A} = \eta {\bf k}  \exp ( \eta k_\mu x^\mu) {\bf A}^{\rm const.} 
= \eta {\bf k}{\bf A},                                                     \eqno (3.30)
$$
so
$$ 
{\bf DF} = {\bf D}^2  {\bf A}  = \partial^\mu\partial_\mu {\bf A} =k^\mu \eta k_\mu \eta \,{\bf A}
= (\eta)^2 {\bf k}^2 {\bf A}   = -{\bf k}^2\, {\bf A} .                   \eqno (3.31)
$$ 
consistent with ${\bf k}^2=0$ and the radiative field
condition ${\bf F}^2 = - {\bf k}{\bf A} {\bf k} {\bf A} = {\bf k}^2{\bf A}^2 = 0$ 
if ${\bf k}$ and ${\bf A}$ anti-commute. It follows that 
$$ 
{\bf D}^2  {\bf A}  = \partial^\mu\partial_\mu {\bf A} =k^\mu \eta k_\mu \eta = {\bf k}^2\eta^2  = 0 .  \eqno (3.32)
$$ 
provides an alternative, Klein-Gordon, form of the photon wave equation.

Plane wave solutions of ${\bf D}{\bf \Phi}= 0$ are
$$
{\bf \Phi}= \exp (\eta k_\mu x^\mu){\bf \Phi}^c ,                                        \eqno (3.33) 
$$ 
where the ${\bf \Phi}^c$ is independent of the space and time coordinates and 
${\bf k} = \gamma^\mu k_\mu$ is the photon wave structor. 
Given (3.25) and (3.33), the field equation ${\bf DF}= 0$ reduces to 
$$
-\eta {\bf D}{\bf \Phi}= -\eta {\bf E}^\nu \partial_\nu  \exp (\eta k_\mu x^\mu){\bf\Phi}^c 
= {\bf k}{\bf \Phi} =0.                                                                   \eqno (3.34)
$$ 
Defining the photon velocity 3-vector ${\bf u} = \gamma^{0i}u_i,\>i=1,2,3$ with ${\bf u}^2 = -1$, 
so that ${\bf k}^2 = (k_0)^2 (\gamma^0 +{\bf u})^2 = 0$.
For photons moving in the $y$-direction, ${\bf u}\rightarrow u_2\gamma^2$, and
(3.34) becomes 
$$ 
(\gamma^0  - u_2\gamma^{2})  {\bf \Phi}^c = 0 \>\>{\rm or,\>equivalently,\>\>}
\gamma^{02} {\bf \Phi}^c = u_2 {\bf \Phi}^c,                                      \eqno (3.35)
$$ 
where $u_2=\pm 1$, corresponding to the direction of
the photon velocity, with unit magnitude corresponding to
the velocity of light. 
Equation (3.35) relates to unpolarized photons, leaving open
the question of finding elements of $Cl_{3,3}$ that commute 
with $\gamma^{02}$, with eigenvalues that distinguish polarization
and the sign of interactions with charged fermions.  
Polarizations are normally described by the 4-vectors 
$\epsilon^i = \epsilon^i_\mu \gamma^\mu,\> i=1,2$, 
orthogonal to the wave-vector ${\bf k} = k_\mu \gamma^\mu$, 
giving  
$$
{\bf k}\>\epsilon^i + \epsilon^i \>{\bf k} = 2 k^\mu \epsilon^i_\mu = 0.   \eqno (3.36)
$$
In the algebraic formulation plane polarizations 
could be described by the eigenvalues of $\gamma^{31}$.

\beginsection  \S4. Description of leptons in terms of the eigenvalues of commuting elements

In the Dirac theory, 4$\times$4  matrices $\bar\gamma$ act 
on 4-component spinors. The Dirac expressions for these matrices (see Table A3), which
can be interpreted as describing Minkowski coordinates in the fermion rest frame, 
are denoted by $\bar\gamma^{*\mu}$ in this work, where the star indicates that 
they are invariant under Lorentz transformations.
The electron/positron distinction is determined by the eigenvalues of 
$\bar{\gamma}^{*0} = -{\bf I}\otimes {\bf R}$,
which are $+1$ for electrons and $-1$ for positrons.
The up/down spin distinction is determined by the eigenvalues
$\pm i$ of $\bar{\gamma}^{*12} = i{\bf R}\otimes {\bf I}$, 
which commutes with $\bar{\gamma}^{*0}$. Hence the
binary eigenvalues of two commuting elements of the $Cl_{1,3}$ algebra
distinguish four states of a lepton. 

The 8$\times$8 representation matrices of $Cl_{3,3}$ act on 
8-component column matrices. These 
components will be shown to distinguish the four
states the two leptons in a given 
generation, and  relate them to commuting elements of $Cl_{3,3}$.
As the squared elements of Clifford algebras are all $\bf \pm {\bf 1}_3$, 
their eigenvalues are necessarily twofold, i.e. $\pm 1$ or $\pm i$,
so that three commuting elements of  $Cl_{3,3}$ are
required to distinguish $2^3=8$ lepton states.
These three elements, and their eigenvalues, 
will be called {\it primary}. The anti-lepton that corresponds to a given
lepton has opposite signs of {\it all} its primary eigenvalues.  
Pair products of the three primary commuting
elements determine three {\it secondary} eigenvalues, while the product of all
three gives a fourth primary eigenvalue, which
determines the direction of time and distinguishes 
fermions from anti-fermions.  Secondary eigenvalues 
have the same values for a lepton and its corresponding anti-lepton.

Let $\gamma^A$, $\gamma^B$ and $\gamma^C$ be commuting Hermitian matrices, 
with eigenvalues $\mu_A=\pm 1$, $\mu_B=\pm 1$ and $\mu_C=\pm 1$.
Each matrix defines a projection operator, e.g.
 ${\bf P}(\mu_A) = {1\over 2}({\bf 1}_3 + \mu_A\gamma^A)$. 
These matrices will be related to elements 
$\gamma$ of $Cl_{3,3}$ where $\gamma$ is time-like, or
$i\gamma$ when $\gamma$ is space-like.  In the following analysis it 
will be assumed that the $\gamma$-matrices defined in Appendix A
refer to the Minkowski coordinates in the lepton rest frame, and
the `star' notation will be employed to distinguish them from 
representation matrices corresponding to unit space-time displacements
in arbitrary frames. 
The aim is to identify  $\gamma^{A}$, $\gamma^{B}$ and $\gamma^{C}$ with 
specific elements of $Cl_{3,3}$. The eight distinct lepton states are  
projected out of an 8-component column matrix by 
$$
{\bf P}(\mu_A, \mu_B, \mu_C) = {\bf P}(\mu_A){\bf P}(\mu_B){\bf P}(\mu_C) 
={1\over 8}({\bf 1}_3+\mu_A \gamma^A)({\bf 1}_3+\mu_B\gamma^B)({\bf 1}_3+\mu_C \gamma^C).  \eqno (4.1)
$$
The space-like anti-commuting elements ${\hat\gamma}^{*12},{\hat\gamma}^{*23}, {\hat\gamma}^{*31}$, 
generate the Lie algebra $SU(2)_{spin}$. $\gamma^A$
can be identified as $i$ times any normalised linear combination of them,
corresponding to the (arbitrary) choice of spin orientation, 
but, as the eigenvalue $\mu_A$ provides no information about this orientation,
it can be assumed that $\gamma^A = i\gamma^{*31}$.

In order that each of the eight eigenstates
corresponds to a single non-zero entry in the column matrix it
is necessary to choose a representation in which all three
matrices $\gamma^B$ and $\gamma^C$ and $\gamma^A = \gamma^{*31}$ are diagonal. 
This is achieved by redefining the $\gamma$-matrix representation using
the similarity transformation $ \hat{\gamma}= {\cal Z} \gamma {\cal Z}^{-1}$, 
defined in Appendix A, giving the 64 $\hat{\gamma}$ matrices 
in Table A2. Another important result of introducing
the  $\hat{\gamma}$ representation is that it block
diagonalises Lorentz transformations and, consequently,
all the matrices that describe structors, as shown in Appendix B. 

The space-like anti-commuting matrices 
${\hat\gamma}^{\pi 6} (= -{\hat\gamma}^{78}),\>{\hat\gamma}^{\pi 7},\>{\hat\gamma}^{\pi 8}$ 
generate the Lie algebra $SU(2)_{isospin}$.  
As all three commute with ${\hat\gamma}^{*12},\>{\hat\gamma}^{*23}$ and ${\hat\gamma}^{*31}$, 
any one of them, or any normalised linear combination could, in principle, 
be identified with $-i\gamma^C$. In practice,
however, $SU(2)_{isospin}$ symmetry is broken so, in the following analysis, 
leptons will be described by the eigenvalues of the diagonal matrix $\gamma^C = i{\hat\gamma}^{*\pi 6}$, 
so that $\mu_C = i\mu_{\pi 6} = \pm 1$. (The `isospin' 
label introduced here provides the same quantum number as the isospin currently employed 
in the description of baryon flavour symmetry.) 

Having identified $\gamma^A$ and $\gamma^C$ with pair products of generators, it 
is clear that $\gamma^B$ could be identified with the time-like matrix ${\hat\gamma}^{*26}$,
but this matrix does not correspond to a readily observable property of
leptons. The alternative is to identify 
$\gamma^B ={\hat\gamma}^{*0} = -{\hat\gamma}^{*26} {\hat\gamma}^{*31} {\hat\gamma}^{\pi 6}$,
which is the time direction in the fermion rest frame. 
The Standard Model was originally formulated when neutrinos 
were thought to have zero mass but, as neutrinos 
and anti-neutrinos are now known to have small non-zero masses, they can be
described by spinors that are eigenstates of $\hat\gamma^{*0}$. 
It follows that $\gamma^B=\hat\gamma^{*0}$, with eigenvalues 
$\mu_B = \mu_{*0} = +1$ for leptons and  $\mu_B = \mu_{*0} = -1$ 
for anti-leptons, giving the lepton state 
identifications summarized in Table 4.1. This table also shows that  
the same quantum numbers can be associated with stable baryons, i.e. neutrons($n$) and protons($p$).
$$\vbox {\settabs 3\columns
	\+ Table 4.1: Lepton identification\cr
	\+||||||||||||||||||||||||||||||||||||||\cr
	\+ & $\mu_B=\mu_{*0}= +1$ & $\mu_B=\mu_{*0} = -1$ \cr
	\+||||||||||||||||||||||||||||||||||||||\cr
	\+$\>\>\>\mu_C=i\mu_{\pi 6} = +1$&$\>\>\>\>\>e^-,\>\> p^-$&$\>\>\>\>\>\bar\nu\!,\>\>n$\cr\+&&\cr
	\+$\>\>\>\mu_C=i\mu_{\pi 6} = -1$&$\>\>\>\>\>\nu\!,\>\>\bar n$&$\>\>\>\>\>e^+, \>\>p^+$\cr
	\+||||||||||||||||||||||||||||||||||||||\cr}
$$
A complete description of lepton states, including 
the spin degree of freedom, is given in 
Table 4.2, which shows the $Cl_{3,3}$ 
algebra to be consistent with neutrinos being described by 
Dirac (4-component) spinors, rather than
2-component spinors. Lepton charges, are given by
$$
\mu_{ Q} = -{1\over 2}\,(\mu_{*0}+i\mu_{\pi 6}) = -{1\over 2}\,(\mu_B+\mu_C), \eqno (4.2)
$$
times the magnitude of the electronic charge $e$. The primary eigenvalues 
$ i\mu_{\pi 6},\>i\mu_{*31},\>\mu_{*0},\>\mu_{*26}$ (in the first four columns of Table 4.2)
have opposite signs for leptons and their corresponding anti-leptons.

$$\vbox {\settabs 8 \columns \+ Table 4.2: Lepton quantum numbers and
	charges \cr \+||||||||||||||||||||||||||||||||||||||||||||\cr
	\+ isospin            &spin             & proper time & & mass/energy &helicity&charge& lepton\cr 
	\+$ C: i\mu_{\pi 6}$ & $ A: i\mu_{*31}$ &$B: \mu_{*0}$ &$ABC:\mu_{*26}$& $BC:
	i\mu_{*\pi60}$&$AB: i\mu_{\pi 2}$&$\>\>\>\>{\mu_Q} $ & state \cr
	\+||||||||||||||||||||||||||||||||||||||||||||\cr
	\+$\>\>\>1$&$\>\>\>1$&$\>\>\>1$&$\>\>\>1$&$\>\>\>1$&$\>\>\>1$ &$\>\>-1$     &$\>\>\>e^-\!\!\downarrow$ \cr\+&&&\cr
	\+$\>\>\>1$&$-1$     &$\>\>\>1$&$-1$     &$\>\>\>1$&$-1$      &$\>\>-1$     &$\>\>\>e^-\!\!\uparrow$\cr\+&&&\cr
	\+$\>\>\>1$&$\>\>\>1$&$-1$     &$-1$     &$-1$     &$-1$      &$\>\>\>\>\>0$&$\>\>\>\>\bar\nu\!\!\downarrow$\cr  \+&&&\cr
	\+$\>\>\>1$&$-1$     &$-1$     &$\>\>\>1$&$-1$     &$\>\>\>1$ &$\>\>\>\>\>0$&$\>\>\>\>\bar\nu \!\!\uparrow$\cr
	\+||||||||||||||||||||||||||||||||||||||||||||\cr
	\+$-1$     &$\>\>\>1$&$\>\>\>1$&$-1$     &$-1$     &$\>\>\>1$ &$\>\>\>\>\>0$&$\>\>\>\>\nu\!\!\downarrow$\cr\+&&&\cr
	\+$-1$     &$-1$     &$\>\>\>1$&$\>\>\>1$&$-1$     &$-1$      &$\>\>\>\>\>0$&$\>\>\>\>\nu\!\!\uparrow$\cr\+&&&\cr
	\+$-1$     &$\>\>\>1$&$-1$     &$\>\>\>1$&$\>\>\>1$&$-1$      &$\>\>\>\>\>1$&$\>\>\>e^+\!\!\downarrow$\cr \+&&&\cr
	\+$-1$     &$-1$     &$-1$     &$-1$     &$\>\>\>1$&$\>\>\>1$ &$\>\>\>\>\>1$&$\>\>\>e^+\!\!\uparrow$\cr
	\+||||||||||||||||||||||||||||||||||||||||||||\cr}
$$
If the leptons states are labelled in the same order as in the last column of Table 4.2, 
entries in the first four columns determine the diagonal matrices 
that correspond to the primary eigenvalues, viz. 
$$\eqalign{
  \gamma^A= i\hat{\gamma}^{*31} =\>&-{\bf R}\otimes {\bf I}\otimes{\bf I} = diag(1 \bar 1  1 \bar 1; 1 \bar 1 1 \bar 1 ),\cr
\>\gamma^B=\hat{\gamma}^{*0}=\>&-{\bf I}\otimes {\bf R}\otimes{\bf I} = diag(1  1  \bar 1  \bar 1;  1  1 \bar 1 \bar 1 ),\cr
\>\gamma^C=i\hat{\gamma}^{\pi 6}=\>&-{\bf I}\otimes {\bf I}\otimes{\bf R}=diag(1  1  1  1;\bar 1 \bar 1  \bar1 \bar 1 ), \cr
\>\gamma^{ABC}=\>& -{\bf R}\otimes {\bf R}\otimes{\bf R} =diag(1 \bar 1\bar 1  1; \bar 1  1  1 \bar 1 ) ,} 
\eqno (4.3)
$$ 
where $\bar 1 \equiv -1$. The structor corresponding 
to $\hat\gamma^{*31}$ is 
$$
{\bf s}(-)= \hat{\gamma}^{\mu \nu}s_{\mu\nu},\{\mu,
\nu =0,1,2,3\},                                                                           \eqno (4.4)
$$
with values of the coefficients $s_{\mu\nu}$ determined by the reference frame.
The structor with eigenvalues corresponding to lepton charges is 
$$\eqalign{
{\cal Q}  =& -{1\over 2}\,(\hat\gamma^{*0}+i\hat\gamma^{\pi 6}) 
= {1\over 2}({\bf I}\otimes {\bf R}\otimes {\bf I}+{\bf I}\otimes {\bf I}\otimes {\bf R})\cr
\equiv &\>{1\over 2}\bigl(diag(\bar1\> \bar1 \>1\> 1;\>\bar1\>\bar1\>1\>1)+ diag(\bar1\>\bar1\>\bar1\>\bar1;\>1\>1\>1\>1)\bigr)= \>diag(\bar1\>\bar1\>0\>0;\>0\>0\>1\>1).}\eqno (4.5)
$$
Its square
$$
{\cal Q}^2  = {1\over 2}\,({\bf 1}_3 + \hat{\gamma}^{*0}i\hat\gamma^{\pi 6}) = diag(1\>1\>0\>0;\>0\>0\>1\>1)
 \eqno (4.6)
$$
has eigenvalues $+1$ for electrons and positrons, 
and zero for neutrinos and anti-neutrinos, giving the mass formula
$$\eqalign{
{\bf M}=& m_\nu {\bf 1}_3+ (m_{\rm e}- m_\nu){\cal Q}^2 \cr
=&{1\over2}m_\nu ({\bf 1}_3 - i\hat\gamma^{\pi 6}\hat{\gamma}^{*0})
+ {1\over2}m_{\rm e}({\bf 1}_3 + i\hat\gamma^{\pi 6} \hat{\gamma}^{*0})\cr
=&diag(m_{\rm e},\>m_{\rm e},\> m_\nu,\> m_\nu ,\>m_\nu ,\>m_\nu ,\>m_{\rm e},\>m_{\rm e})},   \eqno (4.7) 
$$
and ${\bf p}={\bf M}\hat{\gamma}^{*0}$.

In the Standard Model the spin quantum number $s_z$
is related to the helicity quantum number for electrons with momentum 
$\vec p$ defined by $h={ \vec s.\vec p}/p$, where $\vec s$ is the spin 3-vector, 
$\vec p$ is the momentum 3-vector and $p^2 = {\vec p}^{\,2}$ 
(e.g. [21] page 105). With this definition, helicity is found to be   
conserved in high energy interactions, although it is clearly not invariant
under Lorentz transformations that change the sign of $\vec p$. 
In the $Cl_{3,3}$ formalism, the spin quantum number is associated with  
$\hat{\gamma}^{*31}$, which is a component of
the structor ${\bf s}(-) = \hat{\gamma}^{\mu\nu} s_{\mu\nu}$. 
Helicity is identified with the pseudo-vector
$\hat{\gamma}^{*\pi2} = -\hat{\gamma}^{*0}\hat{\gamma}^{*31}$,
which is a component of the structor ${\bf h}(-) = \hat{\gamma}^{\pi\mu} h_\mu$. This 
Lorentz invariant redefinition of helicity is important in the 
analysis of experimental results.

Discrete geometrical transformations of the space-time coordinates 
were given in \S3. It is assumed, in the Standard Model, that
quantum mechanical equivalents can be obtained by expressing these 
transformations in terms of the Dirac algebra, but there is experimental
evidence that particle interactions are not always invariant 
under these transformations, suggesting a need to reformulate
them in terms of the $Cl_{3,3}$ algebra.
Geometrical symmetries are related to the properties of elementary fermions
by replacing the ${\bf E}^\mu$ with their matrix representations $\hat{\gamma}^\mu$. 
Inversion of the spatial coordinates 
corresponds to changing their parity ${\bf \hat P}$, defined by the transformation
$$
{\bf \hat P}: \hat{\gamma}^{\mu}\rightarrow \hat{\gamma}^{0}\hat{\gamma}^{\mu}
(\hat{\gamma}^{0})^{-1} = \hat{\gamma}_{\mu},                            \eqno (4.8)
$$
where $\hat{\gamma}^{0}=\hat{\gamma}_{0}$ is the observer's time direction.
As each coordinate frame, and each fermion, defines its own
time direction, ${\bf \hat P}$ is not invariant in fermion
interactions. The assignment of positive  
parity to fermions and negative parity to anti-fermions, made in the Standard Model,
relates to the time direction $\hat{\gamma}^{*0}$ in the fermion rest frame, rather than
the time direction $\hat{\gamma}^{0}$ in the observer's frame. This is consistent with 
(4.8) if the corresponding Lorentz invariant operator $\cal P$, defined by
$$
{\cal P}: \hat{\gamma}^{*\mu}\rightarrow \hat{\gamma}^{*0}\hat{\gamma}^{*\mu}
(\hat{\gamma}^{*0})^{-1} = \hat{\gamma}_{*\mu},                          \eqno (4.9)
$$
where $\hat{\gamma}^{*0}=(\hat{\gamma}^{*0})^{-1} $, 
is the (Lorentz invariant) proper time, so that 
the reversed spatial coordinates $\hat{\gamma}^{*\mu},\> \mu = 1,2,3$
refer to the fermion's rest frame. As each particle has its own rest 
frame, this can be difficult to relate to experimental results.
Nevertheless, it can be expressed in terms of the Lorentz invariant 
$\hat{\gamma}^\pi\equiv \hat{\gamma}^{*\pi}$ which 
satisfies  ${\cal P}\hat{\gamma}^\pi=-\hat{\gamma}^\pi$. 
The association of parity 
with fermion states, {\it assigned} in the Standard Model, can now be
seen as {\it defining} fermion parities in terms of 
the eigenvalues of $\hat{\gamma}^{\pi 6}$. 
Coordinate reflections also change the parity of the coordinate system
as expressed by the sign of the Lorentz invariant $\hat{\gamma}^{\pi}$. 
For example, reflections in the $\hat{\gamma}^{*31}$ plane in the fermion
rest frame, which produce a reversal of the fermion spin direction, are described by 
$$	
{\bf \hat P}_{31}: \hat{\gamma}^{*\mu}\rightarrow \hat{\gamma}^{*\pi2}
\hat{\gamma}^{*\mu}(\hat{\gamma}^{*\pi2})^{-1}
 = \hat{\gamma}^{*\mu},\>{\rm for}\>  \mu = 0,1,3,\>{\rm or}\>
  -\hat{\gamma}^{*\mu} ,\>{\rm for}\> \mu = 2,\>\> {\rm and}\>\> \pi,     \eqno (4.10)
$$
showing that single coordinate reflections change parity.

The time-reversal operator in an arbitrary coordinate frame has 
the representation ${\bf \hat T}=  \hat{\gamma}^{\pi 0}$ which, again, 
is not Lorentz invariant. This geometrical, or unitary, 
form of time-reversal changes the sign of the Hamiltonian, this problem
being overcome in the Standard Model by including the sign reversal 
$i \rightarrow -i$, making the transformation anti-unitary. 
The $Cl_{3,3}$ algebra provides the proper time-reversal operators 
${\cal T }^k =\hat{\gamma}^{*k0}:\>k=6,7,8$ giving, in the fermion rest frame, 
$$ {\cal T }^k:\> \hat{\gamma}^{*\mu} \rightarrow \hat{\gamma}^{*k0} 
\hat{\gamma}^{*\mu} (\hat{\gamma}^{*k0})^{-1} = -\hat{\gamma}_{*\mu},               \eqno (4.11)
$$
where $k=6,7,8,\pi$. If $k = \pi\> {\rm or} \> 6$ the same unitarity problem 
arises. It is, however, avoided by choosing $k = 7\>\> {\rm or}\> \>8$, 
both of which go beyond $Cl_{1,3}$ space-time geometry, 
and provide unitary, Lorentz invariant, forms of time-reversal. 

All seven quantum numbers that can be constructed from $A$, $B$ and $C$
correspond to algebraic invariants, which are structors if they 
involve either $A$ or $B$. A summary of their physical interpretations is given below: 

\vskip5pt

\settabs\+\indent &{\it quantum no.}\quad&structor\quad\quad\quad\quad\quad\quad\quad\quad\quad\quad&macroscopic interpretation\quad\quad&quantum interpretation&\cr        	
\+&Table 4.3 Physical interpretations of the seven algebraic invariants\cr
\+&||||||||||||||||||||||||||||||||||||||||||||\cr
\+&{\it quantum no.}&{\it algebraic invariant/structor}&{\it macroscopic interpretation}&{\it quantum interpretation}\cr
\+&||||||||||||||||||||||||||||||||||||||||||||\cr
\+&$A:\>\mu_{*31 }$&${\bf s}(-)=\hat\gamma^{*31}s_{31}=\hat\gamma^{\mu\nu}s_{\mu \nu}$&intrinsic angular velocity&  spin\cr
\+&$B:\>\mu_{*0}$&$\hat{\gamma}^{*0}=\hat{\gamma}_{\mu} {dx^\mu/dx^{*0}} $&proper time direction&fermion/anti-fermion\cr
\+&&&&distinction\cr
\+&$C:\>\mu_{\pi6}$&${\hat\gamma}^{\pi 6} = {\hat \gamma}^{8}{\hat \gamma}^{7}$&fermion parity&iso-spin, quantum $i$,\cr
\+&&&&lepton substrate\cr
\+&$BC:\>i\mu_{\pi6 *0}$&${\bf p}={\bf M}\hat{\gamma}_{\mu} {dx^\mu/dx^{*0}}$&4-momentum& as macroscopic\cr
\+&$AC:\>\mu_{*026}$&${\bf s}\hat\gamma^{\pi6 }(+)$&magnetic moment& as macroscopic\cr
\+&$AB:\>\mu_{\pi*2}$&${\bf h (-)= s}{\hat\gamma}^{*0} = \hat{\gamma}^{\pi\mu} h_\mu$&&helicity \cr
\+&$ABC:\>\mu_{*26}$&$\hat{\gamma}^{*26}={\bf h}\hat\gamma^{\pi6} = -\hat{\gamma}^{*31}\hat{\gamma}^{*0}\hat{\gamma}^{\pi6}$&spin angular momentum&as macroscopic\cr
\+&||||||||||||||||||||||||||||||||||||||||||||\cr

\vskip5pt

\beginsection 5. Reformulation of the Dirac equation

The established procedure for obtaining the quantum mechanical equations of
motion for free particles from their classical counterparts is to
replace the momentum 3-vector ${\vec p}= (p_1, p_2, p_3)$ by the operator 
$-i\nabla = -i(\partial_1, \partial_2,\partial_3)$ and 
the energy $E$ by the operator $ i\partial_t$. Wave equations 
are then produced from the action of the
relation between mass, momentum and energy on a wave function. In particular,
the Schr\~odinger equation $i\partial_t \phi= (\nabla^2/2m)\phi $, 
where $\phi$ is the wave-function, is obtained from the 
mass/momentum/energy relation for free particles in classical mechanics, 
i.e. $E= -{1\over 2m}{\vec p}^{\, 2} $. Its solution is the wave function
$\phi =\phi_0 \exp(\pm i (\vec p .\vec x - Et)) $ where $\phi_0$ is constant.

The following analysis clarifies the relationship between the 4$\times$4 
${\bar\gamma}$ Dirac matrix representation of $Cl_{1,3}$ and the 
8$\times$8 ${\hat\gamma}^\mu$ space-time matrices 
of $Cl_{3,3}$. In Appendix A explicit comparisons
are made between representations of the ${\bar\gamma}^\mu$
and ${\hat\gamma}^{*\mu}$ fermion rest frame coordinates. 
The star notation, introduced in \S3, distinguishes
the fermion rest-frame from the arbitrary reference frames
employed by observers. The $\bar\gamma^\mu$ always refer
to the fermion rest frame.

Physical space-time coordinates can be represented either 
by the ${\hat\gamma}^\mu$ matrices or by the familiar Dirac 
${\bar\gamma}$ matrices. The following analysis relates these
alternatives, making it clear that the Dirac formulation is
complicated by the two fermions in any doublet are described
by spatial coordinate systems with opposite parity, corresponding 
to the identification $\bar\gamma^\mu \equiv\> ^a\gamma^\mu$ or
$\bar\gamma^\mu \equiv\> ^b\gamma^\mu$.
 
The relativistic energy/momentum conservation equation for free 
particles is ${\bf p}^2 = E^2 - {\vec p}^{\, 2} = m^2$. In terms
of the $\bar\gamma$ algebra this corresponds to the structor 
equation
$$
{\bf p} ={\bar\gamma}^\mu p_\mu = m{\bar\gamma}^{*0},              \eqno (5.1)
$$
where $\bar\gamma^{*0}$ has the eigenvalue $\mu_{*0}=+1$ for electrons and
$\mu_{*0}=-1$ for positrons. The standard replacement 
$p_{\mu} \to i\partial_\mu $ gives
$$
{\bf p} =\bar\gamma_\mu p^\mu \to  i{\bf D}=i\bar\gamma^\mu \partial_\mu,  \eqno (5.2)
$$
and the relativistic free electron wave equation
$$
(i{\bf D} - m{\bar\gamma}^{0}){\phi}=0.                      \eqno (5.3)
$$ 

Dirac's wave-equation $(i{\bf D} - m){\phi}=0$,
which is currently accepted as providing
the correct description of fermion dynamics, omits 
${\bar\gamma}^{0}$ in (5.3).
It was derived by taking the square root of both sides of the relativistic 
free fermion energy/momentum conservation equation ${\bf p}^2 = m^2$, giving
$$
{\bf p} =\bar\gamma^{\mu} p_\mu = m                                \eqno(5.4)
$$
rather than (5.1). However, no linear combination of the Dirac 
matrices $\bar\gamma^{\mu}$, which all have zero trace, can give 
rise to the right hand side of (5.4).  When $\bar\gamma^{0}$ is
omitted from (5.3), as it is in the Dirac equation, 
$\phi$ becomes subject to Lorentz transformations and 
the mass sign problem of the Dirac theory is produced.

The $Cl_{3,3}$ reformulation is 
obtained by replacing the 4$\times$4 matrices $\bar\gamma $ of $Cl_{1,3}$ 
with the 8$\times$8 matrices $\hat\gamma$ of $Cl_{3,3}$, and substituting 
(5.1) with the expression for 
${\bf p}$ given in \S4. In the case of leptons, the mass $m$
is replaced by diagonal matrix ${\bf M}$ defined in (4.7).
The replacement ${\bf p}\to \hat\gamma^{\pi6}{\bf D} $ then 
gives the free lepton wave-equation 
$$
\hat\gamma^{\pi6}{\bf D}{\bf\Psi} = {\hat\gamma}^{*0}{\bf M}{\bf\Psi},         \eqno (5.5)
$$ 
where the matrix ${\bf\Psi}$ describes all eight states of the lepton doublet.
Solutions of (5.5) are based on recognising that 
$$
{\bf D} = \hat\gamma^\mu \partial_\mu = \hat\gamma^{*0} \partial_{*0}, \eqno (5.6)
$$
and that ${\bf\Psi}$ can be written in terms of the observer's or 
lepton coordinate frames, viz.
$$
{\bf\Psi} = {\bf\Psi}_0 \exp (\hat{\gamma}^{\pi 6}p_\mu x^\mu)
= {\bf\Psi}_0 \exp (\hat{\gamma}^{\pi 6}p_{*0} x^{*0}),        \eqno (5.7)      
$$
where ${\bf\Psi}_0$ is a function of the $p_\mu$, and $x^{*0} $ 
is the proper time, i.e. time in the lepton coordinate frame.
As $p_{*0}$ describes lepton masses,
the right and left hand sides of (5.5) are simply alternative
ways of expressing $\hat\gamma^{\pi6}{\bf D}{\bf\Psi}$. 
The matrix $\hat{\gamma}^{\pi 6}$ that appears in the 
exponent is Lorentz invariant, and (as will be argued in \S11) is also invariant 
under space and time translations. It is interpreted as describing the 
lepton {\it substrate}, corresponding to the substrate of the 
electromagnetic field, described by $\eta$ in \S3.                    

Solutions of the Dirac equation are expressed as 4-spinors, which
correspond to columns in the dimensionless matrix ${\bf\Psi}_0 = {\bf M}^{-1} {\bf p}$.
As shown in Appendix B, the matrices $\hat{\gamma}^{\pi 6},\>\hat\gamma^\mu$, 
and structors expressed in terms of them, are block diagonal. In particular, 
$$
{\bf p}  = \left(\matrix{{\bf p}_a&0\cr0&{\bf p}_b\cr}\right),                              \eqno (5.8)
$$
where
$$
{\bf p}_a  =
\left(\matrix{p_0&0&p_2&-p_1-ip_3\cr 0&p_0&-p_1+ip_3&-p_2\cr
	-p_2&p_1+ip_3&-p_0&0\cr	p_1-ip_3&p_2&0&-p_0\cr}\right)
     = {\bf M}_a \left(\matrix{{\bf a}_1&{\bf a}_2&{\bf a}_3&{\bf a}_4\cr}\right)        \eqno (5.8a)
$$
and
$$
{\bf p}_b  =
\left(\matrix{p_0&0&-p_2&-p_1-ip_3\cr 0&p_0&-p_1+ip_3&p_2\cr
        	p_2&p_1+ip_3&-p_0&0\cr p_1-ip_3&-p_2&0&-p_0\cr}\right)
        ={\bf M}_b\left(\matrix{{\bf b}_1&{\bf b}_2&{\bf b}_3&{\bf b}_4\cr}\right).        \eqno(5.8b)
$$
Here ${\bf a}_i,\>{\bf b}_i,\>\{i=1,2,3,4\}$ denote 4-spinor columns in 
$\Psi_0$. The mass matrix ${\bf M}$ 
$$
{\bf M} = \left(\matrix{ {\bf M}_a&0\cr	0& {\bf M}_b\cr} \right)\>\>
{\rm where} 
\>\>{\bf M}_a = \left(\matrix{m_e {\bf I}&0\cr 0& m_\nu{\bf I}\cr} \right),
\>\>{\bf M}_b = \left(\matrix{m_\nu {\bf I}&0\cr 0& m_e {\bf I}\cr} \right).               \eqno (5.9)
$$
The mass factors in (5.8a) and (5.8b) make ${\bf a}_i,\>{\bf b}_i$ dimensionless.

The difference between charged and neutral lepton masses is conventionally
attributed to the Higgs field, which has the algebraic form 
$$
{\cal H } = (m_{\rm e}- m_\nu){\cal Q}^2,                           \eqno (5.10)
$$
where  $m_{\rm e}>> m_\nu$. Table 5.1 compares the labelling of 
the eight 4-spinor solutions shown in (5.8) with 
the four solutions of the Dirac equation. The latter, 
given in [21]\S{\bf4.6.2}, have a similar structure to those
in (5.8), but are not identical. 

The representations of ${\bf p}_a $ and ${\bf p}_b $  make it apparent that 
they relate to different coordinate systems. In particular, ${^a\gamma^2}$ 
and ${^b\gamma^2}$ have opposite signs, as shown in Table A3. with
the consequence that {\it expressions for the 4-spinors ${\bf\Psi }_a$ and 
${\bf\Psi }_b$ relate to coordinate systems with opposite parity}.
Block diagonalisation enables (5.5) to be expressed as two independent equations, viz.
$$
{\bf D}_a {\bf\Psi }_a = {\bf M}_a {\bf\Psi }_a,\>\>{\bf D}_b {\bf\Psi }_b = {\bf M}_b {\bf\Psi }_b,  \eqno (5.11)
$$
where $ {\bf D}_a$ and ${\bf D}_b$ are defined in Appendix B.
As the projection operators ${\bf P}_a$ and ${\bf P}_b$ commute with 
the ${\hat{\gamma}}^{\mu \nu}$ matrices, the components of ${\bf \Psi}_a$
and ${\bf\Psi}_b$ form 4-component column vectors and are not mixed by 
Lorentz transformations. 

$$\vbox {\settabs 9\columns \+ Table 5.1: Comparison of spinor labelling in the Dirac and CU theories\cr
	\+|||||||||||||||||||||||||||||||||||||||||||||\cr
	\+  &$e^-\!\!\downarrow$&$e^-\!\!\uparrow$ & $\bar\nu\!\!\downarrow$ &$\bar\nu \!\!\uparrow$&$\nu\!\!\downarrow$& $\nu\!\!\uparrow$&$\>\>\>e^+\!\!\downarrow$&$\>\>\>e^+\!\!\uparrow$\cr 
	\+|||||||||||||||||||||||||||||||||||||||||||||\cr
	\+mass &$m_e$& $m_e$& $m_\nu$ &$m_\nu$     &$m_\nu$&$m_\nu$&$\>\>\>m_e$ &$\>\>\>m_e $\cr  \+&&&\cr
	\+$ABC$&$111$&${\bar 1}11$&$1{\bar 1}1$&${\bar 1}{\bar 1}1$&$11{\bar 1}$&${\bar 1}1{\bar 1}$ &$\>\>\>1{\bar 1}{\bar 1}$&$\>\>\>{\bar 1}{\bar 1}{\bar 1} $\cr\+&&&\cr
	\+$Cl_{3,3}$&${\bf a}_1$&${\bf a}_2$&${\bf a}_3$&${\bf a}_4$&${\bf b}_1$&${\bf b}_2$&$\>\>\>{\bf b}_3$&$\>\>\>{\bf b}_4$\cr\+&&&\cr
	\+$\mu_{\pi6}$&$i$&$i$&$i$&$i$&$-i$&$-i$&$\>\>\>-i$&$\>\>\>-i$\cr\+&&&\cr
	\+Dirac [21]&$u_1$&$u_2$&&&&&$u_3\equiv v_2$     &$u_4\equiv v_1$\cr
	\+|||||||||||||||||||||||||||||||||||||||||||||\cr
    \+|||||||||||||||||||||||||||||||||||||||||||||\cr}
$$

The free lepton equation (5.5) can be modified to include interactions
with electromagnetic fields simply by adding the field momentum contribution
to the particle momentum, as is done in Lagrangian
theory. For example, the term the electromagnetic
contribution $e \cal Q\bf A$ can be added to the free particle momentum $\bf p$
to produce the generalized, or {\it canonical}, momentum
$$
{\bf p'}= {\bf p} + e \cal Q\bf A .                                 \eqno (5.12)
$$
With this modification, (5.5) becomes
$$
{\bf p'}{\bf\Psi } \rightarrow \hat{\gamma}^{\pi 6}{\bf D}{\bf\Psi } 
= \hat{\gamma}^{\pi 6}{\hat\gamma}^\mu \partial_\mu {\bf\Psi }= ({\bf M}+ e {\bf A}{\cal Q}){\bf\Psi }.\eqno(5.13) 
$$
The factor ${\bf M}+ e {\bf A}{\cal Q}$ can be brought down 
from the exponent by writing 
$$
{\bf\Psi }'= {\bf\Psi }_0\>\exp({\hat\gamma}^{\pi 6}\int { p'}_\mu dx^\mu ),     \eqno (5.14)
$$
where the exponent is  a line integral, with
$$
 {\bf p'}  = \>\hat{\gamma}^{\pi6} \hat\gamma^{\mu} p'_\mu = {\bf M}  +e{\bf A}{\cal Q},\>\>
{\rm where\>\>} {\bf A} =\> A_\mu \hat\gamma^{\mu},\>{\bf M} = {\cal H } + m_\nu {\bf 1}_3\> \>\>
{\rm and \>\>}{\bf A}{\cal Q}=\>-{1\over 2}(A_\mu \hat\gamma^{\mu})(\hat\gamma^{*0}+i\hat\gamma^{\pi 6} ),\eqno (5.15)
$$
reducing the relativistic wave-equation to
$$
{\hat\gamma}^{\pi 6} {\bf D} {\bf\Psi }'= \hat\gamma^{\mu} p_\mu {\bf\Psi }' = {\bf p} {\bf\Psi }'.   \eqno (5.16)
$$
This formulation shows how the algebraic 
description of the physical substrate, modulated by the wave motion, 
can be incorporated into the lepton wave-equation. It should be possible to 
incorporate interactions with other gauge fields in the same way, 
but this remains to be investigated.

\beginsection \S6.  Reformulation of the electro-weak interaction

$\bf \Psi$, defined in \S5, describes all eight lepton states, 
labelled ${\bf a}_i,\>{\bf b}_i,\>\{i=1,2,3,4\}$, and defined 
in Table 5.1. It follows that the $Cl_{3,3}$ algebra must
contain a description of the weak interaction that 
couples electron and neutrino states. Its Standard Model form is 
$$
{\bf X_\mu}(W) =  {g_W\over 2}{\bf W}_\mu = i{g_W\over 2} ( \sigma_1 W^1_\mu +\sigma_2
W^2_\mu + \sigma_3 W^3_\mu),                                                    \eqno (6.1)
$$
where $g_W$ is the (real) coupling coefficient of leptons to the
weak field potential, $\sigma_{\rm k}$ are the Pauli matrices 
(see Appendix A), and the ${W}^{({\rm k})}_\mu\{\rm k=1,2,3\}$ are 4-vector potential
functions. 

The $Cl_{3,3}$ reformulation is obtained by replacing
$$
i\sigma_1 = i{\bf Q} \rightarrow \gamma^{(1)},\>i\sigma_2 = {\bf P}\rightarrow
\gamma^{(2)},\>i\sigma_3 = -i{\bf R}\rightarrow\gamma^{(3)},                     \eqno (6.2)
$$
where $\gamma^{(1)},\> \gamma^{(2)},\>\gamma^{(3)}$ are
anti-commuting elements of $Cl_{3,3}$ that satisfy 
$\gamma^{(1)} \gamma^{(2)}=\gamma^{(3)}$ and 
$(\gamma^{(1)})^2 =(\gamma^{(2)})^2 =(\gamma^{(3)})^2 = -{\bf 1}_3 $. 
As $\hat\gamma^{\pi6}$ takes eigenvalues for all lepton states
it must correspond to the diagonal Pauli matrix $\sigma_3$, giving
$$
\gamma^{(3)} \equiv -\hat\gamma^{\pi6} = -i{\bf I}\otimes {\bf I}\otimes {\bf R} .    \eqno (6.3)
$$
$\gamma^{(1)},\> \gamma^{(2)},\>\gamma^{(3)}$ are generators
of SU(2) and must satisfy the Coleman-Mandula condition that 
they commute with the matrices that define the physical 
coordinate frame. The SM choice corresponds to identifying
$$
 \gamma^{(1)}\equiv i\hat\gamma^{28} =i{\bf I}\otimes {\bf I}\otimes {\bf Q},
 \>\>\gamma^{(2)} \equiv \hat\gamma^{27}= -{\bf I}\otimes {\bf I}\otimes {\bf P}.    \eqno (6.4)
$$
These matrices do not commute with $\hat\gamma^{2}$, but they do commute 
with the physical coordinate frames $^a\gamma^\mu$ and $^b\gamma^\mu$.
Given that the physical coordinate frame for all fermions is described by 
$\hat\gamma^\mu $, the Coleman-Mandula condition requires
$$
\gamma^{(1)}\equiv i\hat\gamma^{\pi8} ={\bf R}\otimes {\bf R}\otimes {\bf Q},
\>\>\gamma^{(2)} \equiv \hat\gamma^{\pi7}= {\bf R}\otimes {\bf R}\otimes {\bf P}.    \eqno (6.5)
$$
This gives raising and lowering operators that describe charged weak bosons as
$$\eqalign{
	\hat\gamma^+ = \>&\>{-1\over  2}\,(\hat\gamma^{\pi7}+ i\hat\gamma^{\pi8}) 
	= \>\>\>{1\over 2}{\bf R}\otimes {\bf R}\otimes ({\bf P}+ {\bf Q})
	=\left(\matrix{ 0&0\cr {\bf R}\otimes {\bf R}&0\cr }\right),\cr
	\hat\gamma^- = \>&\>{1\over 2}\,(\hat\gamma^{\pi7}- i\hat\gamma^{\pi8})
	= \>\>\>{1\over 2}{\bf R}\otimes {\bf R}\otimes ({\bf Q}-{\bf P})
	=\left(\matrix{ 0& {\bf R}\otimes {\bf R}\cr 0&0\cr }\right).\cr }              \eqno (6.6)
$$
These operators satisfy
$$\eqalign{\hat\gamma^- \hat\gamma^- = 0,\>&\> \hat\gamma^+ \hat\gamma^+ = 0,\cr
	\hat\gamma^-\hat\gamma^+ + \hat\gamma^+\hat\gamma^- = {\bf 1}_3,\>&
	\>\hat\gamma^-\hat\gamma^+ -\hat\gamma^+ \hat\gamma^- = i\hat\gamma^{\pi6},\cr
	\>\hat\gamma^-\hat\gamma^{\pi6} + \hat\gamma^{\pi6}\gamma^- =0,\>&
	\> \hat\gamma^+\hat\gamma^{\pi6} +\hat\gamma^{\pi6}\hat\gamma^+ = 0.}           \eqno (6.7)
$$
Defining 
$ W^+_\kappa = W^1_\kappa - iW^2_\kappa ,\>\>\> W^-_\kappa = W^1_\kappa +iW^2_\kappa $, 
the weak potential can be expressed as
$$
{\bf W} = \hat\gamma^\kappa{\bf W_\kappa}={\bf W}^+\hat\gamma^{+}+ {\bf W}^- \hat\gamma^{-} +
{\bf W}^3 \hat\gamma^{\pi6},                                                              \eqno(6.8)
$$
giving the weak interaction
$$
{\bf X_\mu}(W) =  {g_W\over 2}{\bf W}_\mu = {g_W\over 2} ( \hat\gamma^{28}W^1_\mu +\hat\gamma^{27}
W^2_\mu + \hat\gamma^{\pi6} W^3_\mu)= {g_W\over 2} ( \hat\gamma^+ W^+_\mu +\hat\gamma^{-}
W^-_\mu + \hat\gamma^{\pi6} W^3_\mu).                                                    \eqno (6.9)
$$
The action of $\hat\gamma^\pm $ on ${\bf p} $ gives
$$\hat\gamma^+ {\bf p} =\> \left(\matrix{ 0&0\cr \>{\bf p}_a&0 \cr} \right),\>\>
\hat\gamma^-{\bf p}=\> \left(\matrix{ 0&\>{\bf p}_b\cr 0&0 \cr} \right).                 \eqno (6.10)
$$
The physical interpretation of these equations is that the positively charged 
boson $\hat\gamma^+$ adds a charge to fermions with negative or zero charges,
for example converting $e^-\rightarrow  \nu$ and $\bar\nu\rightarrow  e^+ $;
similarly $\hat\gamma^-$ subtracts a charge, so that $e^+ \rightarrow \bar\nu$
and $\nu\rightarrow  e^- $.  The  ${^a\gamma}^{\mu}$ coordinate frame is
relevant to the top row of ${\bf p} $, while the ${^b\gamma}^{\mu}$ coordinate 
frame is relevant to the bottom row. Hence both $\hat\gamma^+$ and $\hat\gamma^-$
change the parity of the coordinate frame, in agreement with the observed parity 
change produced by the weak interaction. This replaces
the SM explanation of the parity change being a consequence of a `V-A' potential 
produced by {\it chirality}. 
 
The separation of electro-magnetic and weak                                 
interactions is achieved by ensuring that their matrix expressions are linearly independent. 
Following the SM argument this involves the introduction
of a potential ${\bf B}$ giving, in terms of the linearly independent matrices $\hat\gamma^{\pi6}$ 
and $\hat\gamma^{*0}$, the neutral electro-weak component
$$
{\bf X}^3 = {g_W\over 2}\,{\bf W}^3  i\hat\gamma^{\pi6} - {g'\over 2} {\bf B}\hat{\gamma}^{*0}.   \eqno (6.11)
$$
The linearly independent potentials ${\bf B}$ and ${\bf W}^3$ can be expressed as rotations 
through the weak mixing angle $\theta$ of the observable electromagnetic and weak potentials
${\bf A}$ and ${\bf Z}$ , viz.
$$
\left(\matrix{ {\bf W}^3\cr {\bf B}\cr} \right)= \left(\matrix{\cos\theta&\sin\theta\cr -\sin\theta& \cos\theta\cr} \right)\left(\matrix {{\bf Z}\cr{\bf A}\cr} \right) .                                        \eqno (6.12)
$$
Substituting (6.12) into (6.11) gives
$$
{\bf X}^3 = {g_W\over 2}\,({\bf Z}\cos\theta + {\bf A}\sin\theta)  i\hat\gamma^{\pi 6}-
{g'\over 2}(-{\bf Z}\sin\theta + {\bf A}\cos\theta) {\hat\gamma}^{*0}.                      \eqno (6.13)
$$
Comparing coefficients of ${\hat\gamma}^{*0}$ and $i\hat\gamma^{\pi 6}$ in (6.11) with those 
for the electromagnetic interaction, given in (4.5), 
$$
e = g_W\sin\theta = g'\cos\theta, \>\> {\rm and}\>\> \tan\theta = {g'\over g_W}.             \eqno (6.14)
$$
These are the same expressions for the weak mixing angle $\theta$ as are obtained the SM, but do not involve chirality, making {\it the above derivation much simpler than that in the SM} (e.g. see [2], pp.418-421). The neutral component of the weak interaction is therefore
$$
{\bf X}^3 (weak) = {1\over 2}(-g' \sin\theta \>\hat{\gamma}^{*0} +
g_W \cos\theta\> i\hat{\gamma}^{\pi6} ){\bf Z}.                      \eqno(6.15)
$$

\beginsection \S7. Physical interpretation of $Cl_{5,5}(LQ)$

The 32$\times$32 $\Gamma$-matrix representations of the ten anti-commuting 
generators of the lepton/quark algebra $Cl_{5,5}(LQ)$ are constructed by inserting the 
anti-commuting elements $ {\bf I}\otimes{\bf P},\>{\bf P}\otimes{\bf R},\>{\bf I}\otimes{\bf Q},\> {\bf Q}\otimes{\bf R}$ of the  $ Cl_{1,1}(5)\otimes Cl_{1,1}(4)$ algebra in front of the generators of $Cl_{3,3}(L)$ defined in Table A2, to give 
$$\eqalign{
	\Gamma^{\,1}=&\>{\bf I}\otimes{\bf I}\otimes\hat\gamma^1= {\bf I}\otimes{\bf I}\otimes{\bf Q}\otimes{\bf P}\otimes{\bf I}\cr
	\Gamma^{\,2}=&\>{\bf I}\otimes{\bf I}\otimes\hat\gamma^2= -{\bf I}\otimes {\bf I}\otimes{\bf R}\otimes{\bf P}\otimes{\bf R}\rightarrow -{\bf I}\otimes {\bf I}\otimes{\bf P}\cr
\Gamma^{\,3}=&\>{\bf I}\otimes{\bf I}\otimes\hat\gamma^3= -i {\bf I}\otimes{\bf I}\otimes{\bf P}\otimes{\bf P}\otimes{\bf I},\cr
	\Gamma^{\,4}=&\>{\bf I}\otimes{\bf P}\otimes \hat\gamma^6= {\bf I}\otimes{\bf P}\otimes{\bf I}\otimes{\bf Q}\otimes{\bf I}
	\rightarrow {\bf I}\otimes{\bf P}\otimes{\bf Q},\cr
	\Gamma^{\,5}=&\>{\bf P}\otimes{\bf R}\otimes \hat\gamma^6= {\bf P}\otimes{\bf R}\otimes{\bf I}\otimes{\bf Q}\otimes{\bf I}
	\rightarrow {\bf P}\otimes{\bf R}\otimes{\bf Q},\cr
	\Gamma^6=    &\>{\bf R}\otimes{\bf R}\otimes \hat\gamma^6= {\bf R}\otimes{\bf R}\otimes{\bf I}\otimes{\bf Q}\otimes{\bf I}
	\rightarrow {\bf R}\otimes{\bf R}\otimes{\bf Q},\cr
    \Gamma^7=    &\>{\bf I}\otimes{\bf I}\otimes \hat\gamma^{7}= i{\bf I}\otimes{\bf I}\otimes{\bf R}\otimes {\bf P}\otimes{\bf Q},\cr
\Gamma^8=    &\>{\bf I}\otimes{\bf I}\otimes \hat\gamma^{8}= {\bf I}\otimes{\bf I}\otimes{\bf R}\otimes{\bf P}\otimes{\bf P},\cr
\Gamma^{9}=  &\>{\bf I}\otimes{\bf Q}\otimes\hat\gamma^{6} = {\bf I}\otimes{\bf Q}\otimes{\bf I}\otimes{\bf Q}\otimes{\bf I}
\rightarrow {\bf I}\otimes{\bf Q}\otimes{\bf Q},\cr
\Gamma^{10}= &\>{\bf Q}\otimes{\bf R}\otimes\hat\gamma^{6}= {\bf Q}\otimes{\bf R}\otimes{\bf I}\otimes{\bf Q}\otimes{\bf I}
\rightarrow {\bf Q}\otimes {\bf R}\otimes{\bf Q}.}      \eqno (7.1) 
$$
The time direction $\Gamma^{\,0}$ is again defined as the product 
of all the generators, viz.
$$
\Gamma^{\,0} =\Gamma^{\,1}\Gamma^{\,2}\Gamma^{\,3}\Gamma^{\,4}\Gamma^{\,5}
\Gamma^{\,6}\Gamma^{\,7}\Gamma^{\,8}\Gamma^{\,9}\Gamma^{\,10}={\bf I}\otimes{\bf I} \otimes \hat\gamma^0 
=-{\bf I}\otimes{\bf I}\otimes{\bf I}\otimes{\bf R}\otimes{\bf I}\rightarrow -{\bf I}\otimes{\bf I}\otimes{\bf R}.
                                                                  \eqno (7.2)
$$
$\Gamma^4$ and $\Gamma^5$ are not 
observed spatial dimensions, so the space-time volume $\Gamma^\pi$ is 
$$
\Gamma^\pi = \Gamma^0\Gamma^1\Gamma^2\Gamma^3 ={\bf I}\otimes{\bf I}\otimes \hat\gamma^\pi. \eqno(7.3)
$$

The 32$\times$32 matrix representation of $Cl_{5,5}(LQ)$ 
distinguishes the $2^5= 32$ quarks and leptons in the first 
generation in terms of the five binary quantum numbers  $ \mu_A, \mu_B, \mu_C, \mu_D, \mu_E$,
where the first three were defined in \S3.
Their corresponding $\Gamma$ matrices are
$$
\Gamma^A =\Gamma^{31} = {\bf I}\otimes {\bf I}\otimes \hat\gamma^{31},\>\>
\Gamma^B =\Gamma^{0} = {\bf I}\otimes {\bf I}\otimes \hat\gamma^{0},\>\>
\Gamma^C =\Gamma^{\pi6} = \Gamma^{87}= {\bf I}\otimes {\bf I}\otimes \hat\gamma^{\pi6}.\eqno (7.4)
$$

The three factor matrices following $\to$ in (7.1) and (7.2) correspond to  
the first, second and fourth factors in the generator matrices,
and generate a real 8$\times$8 
matrix representation of the $Cl_{3,3}(Q)$ sub-algebra 
of $Cl_{5,5}(LQ)$. Writing the generators of $Cl_{3,3}(Q)$ as 
$\dot{\gamma}$-matrices 
$$\eqalign{
	\dot{\gamma}^2 = -{\bf I}\otimes {\bf I}\otimes{\bf P},
	\> \dot{\gamma}^4 = {\bf I}\otimes {\bf P}\otimes{\bf Q}, 
	\>\dot{\gamma}^5 =&\> {\bf P}\otimes {\bf R}\otimes{\bf Q},\cr
	\dot{\gamma}^6 =\> {\bf R}\otimes {\bf R}\otimes{\bf Q},
	\> \dot{\gamma}^{9} = {\bf I}\otimes {\bf Q}\otimes{\bf Q},
	\> \dot{\gamma}^{10} =&\> {\bf Q}\otimes {\bf R}\otimes{\bf Q}.}\eqno (7.5)
$$

There are two ways to construct additional commuting elements $\Gamma^x, \>\Gamma^y $ 
in the  $ Cl_{1,1}(5)\otimes Cl_{1,1}(4)$ algebra, viz.
$$\eqalign{
({\rm i})\> \dot\gamma^x = &\>{\bf I}\otimes{\bf R}\otimes{\bf I}= ({\bf I}\otimes{\bf P}\otimes{\bf Q})({\bf I}\otimes{\bf Q}\otimes{\bf Q})=\dot\gamma^{49},\cr
\> \dot\gamma^y =& {\bf R}\otimes{\bf I}\otimes{\bf I} =\>({\bf P}\otimes{\bf R}\otimes{\bf Q}) ({\bf Q}\otimes{\bf R}\otimes{\bf Q})=\dot\gamma^{5,10}},\eqno(7.6)
$$
and
$$\eqalign{
({\rm ii})\> \dot\gamma^x = &\> {\bf P}\otimes{\bf Q}\otimes{\bf I}=({\bf I}\otimes{\bf P}\otimes{\bf Q})({\bf P}\otimes{\bf R}\otimes{\bf Q})=\dot\gamma^{45},\cr
\> \dot\gamma^y = &{\bf Q}\otimes{\bf P}\otimes{\bf I} =\>({\bf Q}\otimes{\bf R}\otimes{\bf Q})({\bf I}\otimes{\bf Q}\otimes{\bf Q})=\dot\gamma^{9,10}.}\eqno(7.7)
$$
Model (i) is adopted because it has diagonal matrix representations.
The product of all six generators of $Cl_{3,3}(Q)$ gives the  
time direction $\dot{\gamma}^0 = -{\bf I}\otimes {\bf I}\otimes{\bf R}$
identified in (7.2). The matrices $\Gamma^D$ and $\Gamma^E$ correspond 
to $ \dot\gamma^x$ and $ \dot\gamma^y$ respectively, viz.
$$\eqalign{
	\Gamma^D =\Gamma^4\Gamma^9 =&\>\Gamma^{4,9}={\bf I}\otimes{\bf R}\otimes{\bf I}\otimes{\bf I}\otimes
	{\bf I}\rightarrow {\bf I}\otimes {\bf R}\otimes{\bf I}  , \cr
    \Gamma^E =\Gamma^5\Gamma^{10} =&\>\Gamma^{5,10}={\bf R}\otimes{\bf I}\otimes{\bf I}\otimes{\bf I}\otimes
    {\bf I}\rightarrow {\bf R}\otimes {\bf I}\otimes{\bf I}. }\eqno(7.8) 
$$

Table 7.1 distinguishes leptons and quarks in terms of the new primary
quantum numbers $(\mu_{ D}, \mu_{ E})$ and 
$\mu_{ X}= -\mu_{ D} \mu_{ E}\mu_B $. Fermion charges in this 
table are calculated using 
$$
\mu_Q ={1\over 6}( \mu_ D + \mu_E -\mu_{ D}\mu_E\mu_B) - {1\over 2} \mu_C,         \eqno(7.9)
$$
obtained by replacing $-\mu_B$ in (4.2) with
${1\over 3}( \mu_ D + \mu_E + \mu_X)={1\over 3}( \mu_ D + \mu_E -\mu_{ D}\mu_{ E}\mu_B)$. 

$$\vbox
{\settabs 10 \columns\+ && Table (7.1): Lepton and quark quantum numbers ($\mu_B=1$)\cr 
	\+&&|||||||||||||||||||||||||||\cr 
	\+&&$\>\>\mu_C$&$\>\>\mu_D$&$\>\>\mu_E$&$\>\>\mu_X$&   Q   & fermion &&\cr
	\+&&|||||||||||||||||||||||||||\cr
	\+&&$-1$       &$-1$       &$-1$       &$-1$       &$\>\>0$&$\nu$&\cr  
	\+&&$-1$       &$-1$       &$\>\>1$    &$\>\>1$      &$2/3$  &$\rm u_g$&&\cr 
	\+&&$-1$       &$\>\>1$    &$-1$       &$\>\>1 $     &$2/3$  &$\rm u_r$&&\cr  
	\+&&$-1$       &$\>\>1$    &$\>\>1$    &$-1$      &$2/3$  &$\rm u_b$&&\cr 
	\+&&|||||||||||||||||||||||||||\cr
	\+&&$\>\>1$    &$-1$       &$-1$       &$-1$       &$-1$   &$\rm \>e^-$&&\cr 
	\+&&$\>\>1$    &$-1$       &$\>\>1$    &$\>\>1$      &$-1/3$ &$\rm d_g$&&\cr 
	\+&&$\>\>1$    &$\>\>1$    &$-1$       &$\>\>1$      &$-1/3$ &$\rm d_r$&&\cr  
	\+&&$\>\>1$    &$\>\>1$    &$\>\>1$    &$-1$      &$-1/3$ &$\rm d_b$&&\cr  
	\+&&|||||||||||||||||||||||||||\cr}
$$

The operators $\Gamma^B,\>\Gamma^D,\>\Gamma^E,\>\Gamma^X $ have 
diagonal representations corresponding to the entries in Table 7.1, giving
$$\eqalign{
	\Gamma^A =&\>\>\>\>{\bf I}\otimes{\bf I}\otimes{\bf R}\otimes{\bf I}\otimes{\bf I}
	={\bf 1}_2\otimes {\hat\gamma}^A,\cr	
	\Gamma^C =&\>\>\>\>{\bf I}\otimes{\bf I}\otimes{\bf I}\otimes{\bf I}\otimes{\bf R}
	={\bf 1}_2\otimes {\hat\gamma}^C,\cr
	\Gamma^0 =\Gamma^B=&\>-{\bf I}\otimes{\bf I}\otimes{\bf I}\otimes{\bf R}\otimes{\bf I}\to \dot{\gamma}^0 = -{\bf I}\otimes{\bf I}\otimes{\bf R}\equiv diag(1 1 1 1 ;\bar 1 \bar 1 \bar 1 \bar 1),\cr
	  \Gamma^{D}=&\>\>\>\>{\bf I}\otimes{\bf R}\otimes{\bf I}\otimes{\bf I}\otimes{\bf I}\to \dot\gamma^x ={\bf I}\otimes{\bf R}\otimes{\bf I}\equiv diag(\bar1 \bar 1  1 1;\bar 1 \bar 1 1 1),\cr
	 \Gamma^E =&\>\>{\bf R}\otimes{\bf I}\otimes{\bf I}\otimes{\bf I}\otimes{\bf I}\to \dot\gamma^x 
	 ={\bf R}\otimes{\bf I}\otimes{\bf I}\equiv diag(\bar 1 1 \bar 1 1; \bar 1 1  \bar 1 1),\cr
  -\Gamma^E\Gamma^D\Gamma^B=\Gamma^X=& -{\bf R}\otimes{\bf R}\otimes{\bf I}\otimes{\bf
		R}\otimes{\bf I}\to -{\bf R}\otimes{\bf R}\otimes{\bf R}=-\dot\gamma^x\dot\gamma^y\dot\gamma^0
	\equiv diag(\bar 1  1 1 \bar 1 ; 1 \bar 1 \bar 1 1),  }                                    \eqno (7.10)
$$
where the triple Kronecker products are commuting elements of $Cl_{3,3}(Q)$. 
The charge operator corresponding to (7.9) is 
$$
{\cal Q}= {1\over 6}(\Gamma^X + \Gamma^D + \Gamma^E) - {1\over 2} \Gamma^C. \eqno (7.11)
$$

The standard 3$\times$3 Gell-Mann matrix form of the generators of 
the SU$(3)_{strong}$ group are obtained by deleting
first column and top row in eight of the fifteen 4$\times$4 matrices that 
comprise the generators of the Lie algebra SU(4). 
Expressing these matrices  in terms of ${\bf P},\>{\bf Q},\>{\bf R}$ gives
$$\eqalign{
\bar\lambda_1 =&  \left(\matrix{0&0&0&0\cr 0&0&1&0\cr 0&1&0&0\cr 0&0&0&0\cr}\right) = {1\over 2}({\bf Q}\otimes {\bf Q} -{\bf P}\otimes {\bf P}),\>\>\>
\bar\lambda_2 = \left(\matrix{0&0&0&0\cr 0&0&-i&0\cr 0&i&0&0\cr 0&0&0&0\cr}\right) ={i\over 2}({\bf Q}\otimes {\bf P}-{\bf P}\otimes {\bf Q}),\cr
\bar\lambda_3 = &\left(\matrix{0&0&0&0\cr 0&1&0&0\cr 0&0&-1&0\cr 0&0&0&0\cr}\right) ={1\over 2}({\bf R}\otimes {\bf I} - {\bf I}\otimes {\bf R}),\>\>\>
\bar\lambda_4 = \left(\matrix{0&0&0&0\cr 0&0&0&1\cr 0&0&0&0\cr 0&1&0&0\cr}\right) ={1\over 2}({\bf I + R})\otimes {\bf Q},\cr
\bar\lambda_5 = &\left(\matrix{0&0&0&0\cr 0&0&0&-i\cr 0&0&0&0\cr 0&i&0&0\cr}\right) ={i\over 2}({\bf I + R})\otimes {\bf P},\>\>\>\>\>\>\>\>\>\>\>\>
\bar\lambda_6 = \left(\matrix{0&0&0&0\cr 0&0&0&0\cr 0&0&0&1\cr 0&0&1&0\cr}\right) ={1\over 2}{\bf Q}\otimes ({\bf I+R}) ,\cr
\bar\lambda_7 = &\left(\matrix{0&0&0&0\cr 0&0&0&0\cr 0&0&0&-i\cr 0&0&i&0\cr}\right) ={i\over 2}{\bf P}\otimes ({\bf I+R}) ,\>\>
{\sqrt 3}\,\bar\lambda_8 = \left(\matrix{0&0&0&0\cr 0&1&0&0\cr 0&0&1&0\cr 0&0&0&-2\cr}\right) =-{1\over 2}(2{\bf R}\otimes {\bf R} +{\bf I}\otimes {\bf R} + {\bf R}\otimes {\bf I}).}
                                                                          \eqno (7.12)
$$
The $\bar\lambda_i$ act upon the 4-fermion column matrices shown in Table 7.1, showing 
that the gluons do not interact with leptons. However, gluons do act upon anti-quarks, 
so their algebraic representation as operators that act on both quarks and anti-quarks must 
be expressed in terms of the 8$\times$8 matrices $\lambda_i = \bar{\lambda_i}\otimes {\bf I}, \> i=1,...,8$.
In particular the commuting operators $\lambda_3,\>\lambda_8$ are related to the 
commuting elements of $Cl_{5,5}(LQ)$ and its sub-algebra $Cl_{3,3}(Q)$ by
$$\eqalign{
 2\lambda_3\otimes {\bf I} = &({\bf R}\otimes{\bf I}\otimes{\bf I} - {\bf I}\otimes{\bf R}\otimes{\bf I})
 =\dot\gamma^y - \dot\gamma^x,\cr
 2{\sqrt 3}\>\lambda_8\otimes {\bf I}= &-(2{\bf R}\otimes{\bf R}\otimes{\bf I} + {\bf I}\otimes{\bf R}\otimes{\bf I} +{\bf R}\otimes{\bf I}\otimes{\bf I}  )= -(2\dot\gamma^x\dot\gamma^y +\dot\gamma^x +\dot\gamma^y). }\eqno (7.13)
$$

The model (i)  analysis given above reproduces the known 
properties of quarks and gluons as described by
the Standard Model. It does not introduce the
five dimensional space suggested by the $Cl_{5,5}(LQ)$ algebra.  As 
individual quarks and gluons have never been observed in 3-d space, 
the extra two spatial dimensions must relate to
a gluon substrate that only exists {\it inside} hadrons.  
As  gluons interact strongly within hadrons 
it is reasonable to suppose that they form a 
coherent jelly-like substrate. This is transparent to 
leptons, which have no colour charge.
This model would explain the strength of
long range quark/quark interactions within the jelly 
and why individual quarks are never observed in 3-d space. 
It also suggests that quark/quark interactions 
could be expressed in terms of quark-jelly interactions, 
with the jelly adding effective mass to the quarks.  

\vskip10pt

\beginsection \S8. Physical interpretation of $Cl_{7,7}$

The extension of the ten generators of  $Cl_{5,5}(LQ)$, defined in (7.1),
to the fourteen anti-commuting generators of  $Cl_{7,7}$ follows 
the same pattern used to extend the $Cl_{3,3}(L)$ algebra to $Cl_{5,5}(LQ)$ 
in \S7, viz.
$$\eqalign{
\bar\Gamma^{1}=\>&{\bf I}\otimes{\bf I}\otimes\Gamma^1 ={\bf I}\otimes{\bf I}\otimes{\bf I}\otimes{\bf I}\otimes{\bf Q}\otimes{\bf P}\otimes{\bf I},\>\>\>\>\>\>\>\>\>\>\bar\Gamma^{6}= {\bf R}\otimes{\bf R}\otimes\Gamma^6 ={\bf R}\otimes{\bf R}\otimes{\bf R}\otimes{\bf R}\otimes{\bf I}\otimes{\bf Q}\otimes{\bf I},\cr
\bar\Gamma^{2}=\>&{\bf I}\otimes{\bf I}\otimes\Gamma^2 =-{\bf I}\otimes{\bf I}\otimes{\bf I}\otimes{\bf I}\otimes{\bf R}\otimes{\bf P}\otimes{\bf R},\>\>\>\>\>\>\bar\Gamma^{7}= {\bf I}\>\otimes{\bf I}\otimes\Gamma^7 =i{\bf I}\otimes{\bf I}\otimes{\bf I}\otimes{\bf I}\otimes{\bf R}\otimes{\bf P}\otimes{\bf Q},\cr
\bar\Gamma^{3}= \>&{\bf I}\otimes{\bf I}\otimes\Gamma^3 =-i{\bf I}\otimes{\bf I}\otimes{\bf I}\otimes{\bf I}\otimes{\bf P}\otimes{\bf P}\otimes{\bf I},\>\>\>\>\>\>\bar\Gamma^{8}= {\bf I}\otimes{\bf I}\otimes\Gamma^8 ={\bf I}\otimes{\bf I}\otimes{\bf I}\otimes{\bf I}\otimes{\bf R}\otimes{\bf P}\otimes{\bf P},\cr
\bar\Gamma^{4}= \>&{\bf R}\>\otimes{\bf R}\otimes\Gamma^4 ={\bf R}\otimes{\bf R}\otimes{\bf I}\otimes{\bf P}\otimes{\bf I}\otimes{\bf Q}\otimes{\bf I},\>\>\bar\Gamma^{9}={\bf R}\otimes{\bf R}\otimes\Gamma^9 ={\bf R}\otimes{\bf R}\otimes{\bf I}\otimes{\bf Q}\otimes{\bf I}\>\otimes{\bf Q}\otimes{\bf I},\cr
\bar\Gamma^{5}= \>&{\bf R}\>\otimes{\bf R}\otimes\Gamma^5 ={\bf R}\otimes{\bf R}\otimes{\bf P}\otimes{\bf R}\otimes{\bf I}\otimes{\bf Q}\otimes{\bf I},\>\>\bar\Gamma^{10}= {\bf R}\otimes{\bf R}\otimes\Gamma^{10} ={\bf R}\otimes{\bf R}\otimes{\bf Q}\otimes{\bf R}\otimes{\bf I}\>\otimes{\bf Q}\otimes{\bf I},\cr
&\>\>\>\>\>\>\>\>\>\>\bar\Gamma^{a}= \>{\bf I}\otimes{\bf P}\otimes{\bf I}\otimes{\bf I}\otimes{\bf I}\otimes{\bf Q}\otimes{\bf I},
\>\>\>\>\>\>\>\>\>\>\bar\Gamma^{c}= {\bf I}\otimes{\bf Q}\otimes{\bf I}\otimes{\bf I}\otimes{\bf I}\otimes{\bf Q}\otimes{\bf I},\cr
&\>\>\>\>\>\>\>\>\>\>\bar\Gamma^{b}= \>{\bf P}\otimes{\bf R}\otimes{\bf I}\otimes{\bf I}\otimes{\bf I}\otimes{\bf Q}\otimes{\bf I},\>\>\>\>\>\>\>\>\>\>
\bar\Gamma^{d}= {\bf Q}\otimes{\bf R}\otimes{\bf I}\otimes{\bf I}\otimes{\bf I}\otimes{\bf Q}\otimes{\bf I}
		 .\cr}\eqno (8.1)
$$
\vskip 10pt

The product of all fourteen generators of $Cl_{7,7}$ gives an expression for unit time intervals 
consistent with that previously identified for its sub-algebras $Cl_{3,3}(L)$ and $Cl_{5,5}(LQ)$, i.e.
$$
\bar\Gamma^{0}=\bar\Gamma^{1}\bar\Gamma^{2}...\bar\Gamma^{c}\bar\Gamma^{d}
=-{\bf I}\otimes{\bf I}\otimes{\bf I}\otimes{\bf I}\otimes{\bf I}\otimes{\bf R}\otimes{\bf I}
={\bf 1 }_2\otimes {\Gamma}^0 ={\bf 1 }_4\otimes \hat{\gamma}^0,                                \eqno (8.2)
$$
and $\bar\Gamma^\pi$ is defined as  
$$
\bar \Gamma^\pi = \bar\Gamma^0\bar\Gamma^1\bar\Gamma^2\bar\Gamma^3 = {\bf I}\otimes{\bf I}\otimes{\bf I}\otimes{\bf I}\otimes \hat{\gamma}^\pi =i{\bf I}\otimes{\bf I}\otimes{\bf I}\otimes{\bf I}\otimes{\bf I}\otimes{\bf Q\otimes{\bf R}} . \eqno(8.3)
$$
\vskip10pt
The five quantum numbers already identified 
in the analysis of the sub-algebra $Cl_{5,5}(LQ)$ 
correspond to the $\bar{\Gamma}$ matrices
$$\eqalign{
\bar\Gamma^{ A}= {\bf I}\otimes{\bf I}\otimes\Gamma^{ A}
=&\>{\bf I}\otimes{\bf I}\otimes{\bf I}\otimes{\bf I}\otimes{\bf R}\otimes{\bf I}\otimes{\bf I}
={\bf 1}_4\otimes \hat{\gamma}^A,\cr	
\bar\Gamma^{ C}= {\bf I}\otimes{\bf I}\otimes\Gamma^{ C}
=&\>{\bf I}\otimes{\bf I}\otimes{\bf I}\otimes{\bf I}\otimes{\bf I}\otimes{\bf I}\otimes{\bf R}
={\bf 1}_4\otimes \gamma^{C},\cr
\bar\Gamma^{B}\equiv \bar\Gamma^{ 0}= {\bf I}\otimes{\bf I}\otimes\Gamma^{ B}
=&-{\bf I}\otimes{\bf I}\otimes{\bf I}\otimes{\bf I}\otimes{\bf I}\otimes{\bf R}\otimes{\bf I}
={\bf 1}_4\otimes\gamma^B,\cr
\bar\Gamma^{ D}= {\bf I}\otimes{\bf I}\otimes\Gamma^{ D}
=&\>-{\bf I}\otimes{\bf I}\otimes{\bf R}\otimes{\bf R}\otimes{\bf I}\otimes{\bf R}\otimes{\bf I}
={\bf 1}_2\otimes{\bf R}\otimes{\bf R}\otimes{\gamma}^B,\cr
\bar\Gamma^{ E}= {\bf I}\otimes{\bf I}\otimes\Gamma^{E} 
=&{\bf I}\otimes{\bf I}\otimes{\bf R}\otimes{\bf I}\otimes{\bf I}\otimes{\bf R}\otimes{\bf I}
={\bf 1}_2\otimes{\bf R}\otimes{\bf I}\otimes{\gamma}^B. }     \eqno (8.4)
$$

The $Cl_{3,3}(G)$ sub-algebra has generators defined by the first two and seventh 
factors of the corresponding $Cl_{7,7}$ generators, i.e. 
$$\eqalign{
	\ddot{\gamma}^{a}=& {\bf I}\otimes{\bf P}\otimes{\bf Q},\>\ddot{\gamma}^{b}= {\bf P}\otimes{\bf R}\otimes{\bf Q},\>
	\ddot{\gamma}^{c}={\bf I}\otimes{\bf Q}\otimes{\bf Q},\cr\>\ddot{\gamma}^{d}= 
	&{\bf Q}\otimes{\bf R}\otimes{\bf Q},\>\ddot{\gamma}^{2}
	= {\bf I}\otimes{\bf I}\otimes{\bf P},\>\ddot{\gamma}^{6}
	= {\bf R}\otimes{\bf R}\otimes{\bf Q}.\cr }                      \eqno(8.5)
$$

The commuting elements of $Cl_{3,3}(G)$ are the diagonal matrices
$$
\ddot\gamma^F =\ddot{\gamma}^{ac}= {\bf I}\otimes{\bf R}\otimes{\bf I},\>\>
\ddot\gamma^G =\ddot{\gamma}^{bd}={\bf R}\otimes{\bf I}\otimes{\bf I},\>
\ddot\gamma^H = -{\bf R}\otimes{\bf R}\otimes{\bf R},      \eqno (8.6)
$$
where $\ddot\gamma^H = - \ddot\gamma^F \ddot\gamma^G \ddot\gamma^C$. These determine
the remaining two commuting elements of $Cl_{7,7}$, viz.
$$
\bar\Gamma^F= {\bf I}\otimes{\bf R}\otimes{\bf I}\otimes{\bf I}\otimes{\bf I}\otimes{\bf I}\otimes{\bf I},\>
\bar\Gamma^G= {\bf R}\otimes{\bf I}\otimes{\bf I}\otimes{\bf I}\otimes{\bf I}\otimes{\bf I}\otimes
{\bf I}.                                                                            \eqno(8.7)
$$
The above equations are almost identical to those for the $\dot\gamma$ 
matrices given in \S7, showing the description of generations to
have the same pattern as that of leptons and quarks. with 
correspondences F$\leftrightarrow$D, G$\leftrightarrow$E,
C$\leftrightarrow$B, H$\leftrightarrow$X.

The quantum numbers used to construct Tables 8.1 and 8.2 are
$\mu_F =\mu_{ac}$, $\mu_G= \mu_{bd} $ and $\mu_{H}= -\mu_C\mu_F\mu_G $.
As corresponding anti-fermions have opposite signs of all these quantum numbers, 
they are omitted from these tables. 
A single expression for the charges on the first three (observed) generations
is only obtained if the $\mu_F$ and $\mu_G$ quantum numbers are parity dependent.
This means that the algebraic structure of the weak interaction that was derived in
\S6 needs further elaboration, but this will not be followed up in this work.
 
$$\vbox
{\settabs 10 \columns\+ &&  Table (8.1): Quantum numbers for lepton generations ($\mu_B =1$) \cr 
	\+&&|||||||||||||||||||||||||||\cr 
	\+&&$\>\>\mu_C$&$\>\>\mu_F$&$\>\>\mu_G$&$\>\>\mu_H$&   Q   & lepton&&\cr
	\+&&|||||||||||||||||||||||||||\cr
	\+&&$-1$       &$\>\>1$    &$\>\>1$    &$\>\>1$    &$-2$   &$l^-_2$&&\cr  
	\+&&$-1$       &$-1$       &$-1$       &$\>\>1$    &$\>\>0$&$\nu_e$&&\cr 
	\+&&$-1$       &$-1$       &$\>\>1$    &$-1 $      &$\>\>0$&$\nu_\mu$&&\cr  
	\+&&$-1$       &$\>\>1$    & $-1$      &$-1$       &$\>\>0$&$\nu_\tau$&&\cr 
	\+&&|||||||||||||||||||||||||||\cr
	\+&&$\>\>1$    &$-1$       &$-1$       &$-1$       &$\>\>1$&$l^+_1$&&\cr 
	\+&&$\>\>1$    &$\>\>1$    &$\>\>1$    &$-1$       &$-1$   &$\rm \>e^-$&&\cr 
	\+&&$\>\>1$    &$\>\>1$    &$-1$       &$\>\>1$    &$-1$   &$\mu^-$&&\cr  
	\+&&$\>\>1$    &$-1$       &$\>\>1$    &$\>\>1$    &$-1$   &$\tau^-$&&\cr  
	\+&&|||||||||||||||||||||||||||\cr}
$$

\vskip 5pt

$$\vbox
{\settabs 10 \columns\+ && Table (8.2):Quantum numbers for b quark generations ($\mu_B =1$) \cr 
	\+&&|||||||||||||||||||||||||||\cr 
	\+&&$\>\>\mu_C$&$\>\>\mu_F$&$\>\>\mu_G$&$\>\>\mu_H$&   Q   & fermion&&\cr
	\+&&|||||||||||||||||||||||||||\cr
	\+&&$-1$       &$\>\>1$    &$\>\>1$    &$\>\>1$    &$-4/3$ &$q^{-4/3}$&\cr  
	\+&&$-1$       &$-1$       &$-1$       &$\>\>1$    &$2/3$  &$\rm u$&&\cr 
	\+&&$-1$       &$-1$       &$\>\>1$    &$-1$       &$2/3$  &$\rm c$&&\cr  
	\+&&$-1$       &$\>\>1$    &$-1$       &$-1$       &$2/3$  &$\rm t$&&\cr 
	\+&&|||||||||||||||||||||||||||\cr
	\+&&$\>\>1$    &$-1$       &$-1$       &$\>\>1$    &5/3    &$q^{5/3}$&&\cr 
	\+&&$\>\>1$    &$\>\>1$    &$\>\>1$    &$\>\>1$    &$-1/3$ &$\rm d$&&\cr 
	\+&&$\>\>1$    &$\>\>1$    &$-1$       &$-1$       &$-1/3$ &$\rm s$&&\cr  
	\+&&$\>\>1$    &$-1$       &$\>\>1$    &$-1$       &$-1/3$ &$\rm b$&&\cr  
	\+&&|||||||||||||||||||||||||||\cr}
$$

Electric charges are determined, again in analogy with 
\S7, by substituting the expression 
$(\mu_F+ \mu_G +\mu_H )$ for $\mu_C$ in (7.11), giving the charges 
on all fermions as
$$
\mu_{\cal Q} = {1\over 6}(\mu_X + \mu_D +\mu_E) - {1\over 2}( \mu_H + \mu_F +\mu_G ).\eqno (8.8)
$$
The corresponding charge operator expression is
$$
{\cal Q} = {1\over 6}(\bar{\Gamma}^{ X} + \bar{\Gamma}^{D} +\bar{\Gamma}^{E})
- {1\over 2}( \bar{\Gamma}^{ F}+ \bar{\Gamma}^{ G} +\bar{\Gamma}^{ H} ). \eqno (8.9)
$$
These formulae give the same charges on fermions in all three known generations, as
observed, but predicts different charges on fermions in the predicted, but presently unobserved,
fourth generation. In particular, Table 8.1 shows that fourth 
generation leptons carry either two negative charges or a single positive charge. 
Crucially, this generation has no neutrinos, in accord 
with the experimental evidence that only three types of neutrino exist. 

All four generations have fermion doublets and there is 
good experimental evidence showing that weak interactions relating
the two fermion components of a given doublet are the same for all the 
three known generations, providing the origin of the mass 
differences between their components.
Additional bosons might produce an SU(3)$_{generation}$
gauge field, related to $Cl_{3,3}(G)$ in the same way that 
$SU(3)_{strong}$ is related to $Cl_{3,3}(Q)$. 
The two commuting elements of its Lie algebra are provided
by linear combinations of $\gamma^F,\>\gamma^G$ and $\gamma^H$.
The eight bosons defined by this field would 
be neutral and possibly massive, but
given that the dominant contribution to electron 
mass is due to the Higgs boson,
it is more likely that the masses of the 
second and third generation fermions arise 
from a similar mechanism. A second reason 
for thinking that $SU(3)_{generation}$ bosons
are light is that they interact with neutrinos, 
possibly providing their very small masses. 

Experimental evidence for interactions between 
quarks, other than that produced by gluons, 
is provided by the approximate SU(3)$_{flavour}$ symmetry 
associated with the quark triplet (u, d, s), which provides 
a qualitative explanation of
the baryon and meson mass spectra. As this has already been 
studied in great detail (e.g. see Chapter 9 of [21]) it is only necessary
to relate the existing formalism to the $Cl_{7,7}$ algebra.  
Reference to Table 8.2 
shows that $\mu_F=-\mu_G$ and $\mu_H=\mu_B$ for the four quarks (c, u, d, s),
so that quantum numbers  $\mu_F$, $\mu_H$ and $\mu_C$ are sufficient to
distinguish these quarks and their anti-quarks. 
Quark charges are related to the isospin and 
and hypercharge quantum numbers (given in [1], page 389) using the
Gell-Mann-Nishima formula, viz. ${\mu_Q} = I_3 + Y/2$.
$$\vbox
{\settabs 8 \columns\+  Table 8.3: Quark flavour\cr  
	\+||||||||||||||||||||||||||||||||||||||\cr
	\+$\mu_{F}=-\mu_{G}\>\>\>\>\>\>$&$\mu_{H}=\mu_{B}$&$\>\>\>\>\mu_C$&$\>\>\>I_3$&$\>\>\>Y$&$\>\>\>\mu_{ Q}$&quark\cr
	\+||||||||||||||||||||||||||||||||||||||\cr
	\+$\>\>\>\>\>\>1\>\>$&$\>\>\>\>\>1$&$\>\>-1$ &$\>\>\>0$  &$\>\>4/3$  &$\>\>2/3$&$\>\>\>\>\rm c$\cr
	\+$\>\>\>\>-1\>\>$&$\>\>\>\>\>1$&$\>\>-1$ &$\>\>\>1/2$&$\>\>1/3$&$\>\>2/3$&$\>\>\>\>\rm u$\cr
	\+$\>\>\>\>-1\>\>$&$\>\>\>\>\>1$&$\>\>\>\>\>1$&$-1/2$ &$\>\>1/3$&$-1/3$&$\>\>\>\>\rm d$\cr
	\+$\>\>\>\>\>\>1\>\>$&$\>\>\>\>\>1$&$\>\>\>\>\>1$&$\>\>\>0$&$-2/3$ &$-1/3$  &$\>\>\>\>\rm s$\cr
	\+||||||||||||||||||||||||||||||||||||||\cr
	\+$\>\>\>\>-1\>\>$   &$\>\>-1$&$\>\>\>\>\>1$&$\>\>\>0$&$-4/3$&$-2/3$ &$\>\>\>\>\bar{\rm c}$\cr
	\+$\>\>\>\>\>\>1\>\>$&$\>\>-1$&$\>\>\>\>\>1$&$-1/2$&$-1/3$&$-2/3$ &$\>\>\>\>\bar{\rm u}$\cr  	 
	\+$\>\>\>\>\>\>1\>\>$&$\>\>-1$&$\>\>-1$ &$\>\>\>1/2$&$-1/3$&$\>\>1/3$&$\>\>\>\>\bar{\rm d}$\cr 
	\+$\>\>\>\>-1\>\>$&$\>\>-1$&$\>\>-1$ &$\>\>\>0$&$\>\>\>2/3$&$\>\>1/3$&$\>\>\>\>\bar{\rm s}$\cr 
	\+||||||||||||||||||||||||||||||||||||||\cr}
$$

The algebraic 
relationship between fermions in the first three generations and
fermions in the fourth generation has been shown to be analogous to the relationship
between quarks and leptons. This suggests that this distinction is 
related to wave-function substrates, and that the gauge field that 
produces  mass differences in the first three generations does not act on fourth 
generation fermions. Pressing the analogy further suggests that
large regions of space cannot be occupied by fermions in the first
three generations. Stability, and lack of interactions, makes fourth 
generation fermions possible candidates for producing the constituents 
of dark matter. This accords with the fact that dark matter has 
only been observed through its gravitational effects,
suggesting that it mostly consists of separate, 
electrically neutral, fourth generation composites 
that will be described elsewhere.

\beginsection \S9. General relativity

The algebraic formalism for general relativity is obtained
by generalising the Minkowski coordinates ${\bf E}_\mu$, 
which are the same at all points of space and time, to the Riemannian 
coordinates ${\cal E}_\mu$, which are subject to  continuous variations. 
The generalization of the Clifford algebra to allow 
for the space-time dependence of the ${\cal E}_\mu$ 
was shown in [24] to lead to Einstein's field                       
equations, but this result has not previously 
been related to the analysis in \S3, as is done below. 

The algebraic expression for the Riemannian metric tensor is 
$$
 {\cal E}_\mu {\cal E}_\nu + {\cal E}_\nu {\cal E}_\mu = 2g_{\mu\nu},\eqno (9.1)
$$
with the usual relation between covariant and contravariant
suffices, i.e. ${\cal E}_\nu =g_{\mu\nu} {\cal E}^\mu$. 
As (9.1) is isomorphic to (3.1),
relationships between the $\cal E_\nu$ are
isomorphic to those given \$3 for the $\bf E_\nu$.
For example, following (3.6), the 4-volume element is given by
 $$
 {\cal E}_\pi = {1\over 4!} \epsilon^{\mu \nu \kappa \tau}  {\cal
 	E}_\mu {\cal E}_\nu {\cal E}_\kappa {\cal E}_\tau , \eqno (9.2)
 $$
 so that $({\cal E}_\pi)^2 = g$ is the determinant of the
 4$\times$4 matrix of the $g_{\mu \nu}$.
Defining ${\cal E}^{\nu\kappa}=
{1\over 2}({\cal E}^{\nu}{\cal E}^{\kappa}-{\cal E}^{\kappa}{\cal E}^{\nu})$, 
gives a closure relation isomorphic to (3.10), viz.
$$
{\cal E}^\mu
{\cal E}^{\nu\kappa} = \>\epsilon^{\mu \nu \kappa\tau}{\cal
	E}_{\kappa\tau} + g^{\mu\nu}{\cal E}^\kappa - g^{\mu\kappa}{\cal
	E}^\nu.                                                             \eqno (9.3)
$$

The space-time dependence of the ${\cal E}_\mu $ is given by 
$$
\partial_\kappa {\cal E}_\mu = \Gamma^\tau_{\kappa \mu} {\cal E}_\tau\>,\>\>
\partial_\kappa {\cal E}^\tau = -\Gamma^\tau_{\kappa \mu} {\cal E}^\mu
,\>\partial_\mu {\cal E}_\pi = \Gamma^\kappa_{\kappa \mu} {\cal
	E}_\pi\,                                                             \eqno (9.4)
$$
where $\Gamma^\tau_{\kappa \mu}= {1\over
	2}g^{\tau\lambda}(\partial_\kappa g_{\lambda \mu}+ \partial_\mu
g_{\kappa\lambda}-\partial_\lambda g_{\kappa \mu})$, as usual. Particle
displacements in space-time take the same form as they do
in the Minkowski metric (3.1), i.e.
$$
d{\bf x}=\>{\cal E}_{*0}ds = {\cal  E}_\mu dx^\mu,\> \mu = 0,1,2,3
\>\>{\rm so \>\>that } \>\> {\cal E}_{*0}= {\cal  E}_\mu {dx^\mu\over ds} 
\>\>{\rm and}\>\>(d{\bf x})^2 =({\cal E}_{*0} ds)^2 = (ds)^2.                                                              \eqno(9.5)
$$
In this equation space-time particle displacements are
denoted $ds$, following the standard notation in relativity 
theory, rather than $dx^{*0}$.
The star notation for unit time intervals is the same as that 
used in (3.17), viz. ${\cal E}_{*0}$. Non-interacting 
particles follow geodesic paths that satisfy 
$$\eqalign{
{d{\cal E}_{*0}\over ds}  
=&\> {d\over ds} \bigl({\cal  E}_\mu {dx^\mu\over ds}\bigr) 
= {\cal  E}_\mu{d^2 x^\mu\over ds^2} +  {d {\cal  E}_\mu \over ds}{dx^\mu\over ds} \cr
=&\> {\cal  E}_\mu{d^2 x^\mu\over ds^2} + {dx^\mu\over ds}{dx^\nu\over ds}\Gamma^\tau_{\mu \nu} {\cal  E}_\tau \cr
= &\>{\cal  E}_\tau \bigl({d^2 x^\tau\over ds^2} + {dx^\mu\over ds}{dx^\nu\over ds}\Gamma^\tau_{\mu \nu}\bigr)= 0,}  \eqno (9.6)
$$
where the coefficients of ${\cal  E}_\tau$ provide
the usual tensor expression. 
Differentiating the structor ${\bf A} = A_\mu {\cal
	E}^\mu = A^\nu {\cal E}_\nu $ gives
$$
\partial_\kappa {\bf A} = (A_\mu \partial_\kappa {\cal E}^\mu +
{\cal E}^\mu \partial_\kappa A_\mu) = {\cal E}^\mu (\partial_\kappa
A_\mu + \Gamma^\tau_{\mu \kappa} A_\tau )= {\cal E}^\mu A_{\mu
	;\,\kappa},                                                       \eqno (9.7)
$$
where $A_{\mu;\kappa}$ is
the {\it covariant} differential of $A_\mu$. The  structor form of
(9.7) is produced by the action of the operator 
${\cal D} = {\cal E}^\mu \partial_\mu $ on ${\bf A}$, which 
defines 
$$
{\cal F}={\cal D} {\bf A}  = {\cal E}^{\kappa}{\cal E}^\mu  (\partial_\kappa
A_\mu + \Gamma^\tau_{\kappa\mu}A_\tau) = ({\cal E}^{\kappa\mu} +
g^{\kappa\mu})(\partial_\kappa A_\mu +
\Gamma^\tau_{\kappa\mu}A_\tau)= {\cal E}^{\kappa\mu}\partial_\kappa
A_\mu +  A^\kappa_{\>\>;\kappa} .                                     \eqno (9.8)
$$
If $\bf A$ is interpreted as a potential function, then ${\cal F}=
{\cal D} {\bf A}$ is the corresponding field. This has an invariant
part, $A^\kappa_{\>\>;\kappa}$  and an interactive part ${\cal
	E}^{\kappa\mu}\partial_\kappa A_\mu = {1\over 2}{\cal
	E}^{\kappa\mu}(\partial_\kappa A_\mu - \partial_\mu A_\kappa )$
which couples to the appropriate charge. Maxwell's equations in vacuo
then take the form ${\cal DF}= {\cal D}^2 {\cal A}=0$, if the gauge is chosen so that 
 $A^\kappa_{\>\>;\kappa} = 0$. 

Applying  the differential operator $\partial_\mu$ twice gives 
$$
(\partial_\mu \partial_\nu - \partial_\nu \partial_\mu) {\bf A} =
(\partial_\mu \partial_\nu - \partial_\nu \partial_\mu)A_\kappa
{\cal E}^\kappa  = -
R_{\mu\nu\tau}^{\>\>\>\>\>\>\>\kappa} A_\kappa {\cal E^\tau},\eqno (9.9)
$$
where
$$
R_{\mu\nu\tau}^{\>\>\>\>\>\>\>\kappa}=
\partial_\mu\Gamma^\kappa_{\>\>\tau\nu} - \partial_\nu\Gamma^\kappa_{\>\>\tau\mu}
+ \Gamma^\kappa_{\>\>\sigma\nu}\Gamma^\sigma_{\>\>\tau\mu} -
\Gamma^\kappa_{\>\>\sigma\mu}\Gamma^\sigma_{\>\>\tau\nu}             \eqno (9.10)
$$
is the  Riemann-Christoffel curvature tensor. The
differential operators only commute if  $R^\alpha_{\mu\nu\tau}$
vanishes, i.e. in flat space-time.
In order to obtain the structor equation corresponding to (9.9)
it is necessary to define 
$$
{\cal D}_\wedge = {1\over 2}{\cal E}^{\mu \nu}(\partial_\mu \partial_\nu -
\partial_\nu\partial_\mu).                                              \eqno (9.11)
$$
This gives
$$\eqalign{
{\cal D}_\wedge	{\bf A} 
    = &- {1\over 2}{\cal E}^{\mu \nu}R_{\mu\nu\tau}^{\>\>\>\>\>\>\>\kappa} A_\kappa {\cal E^\tau} \cr 
	=&- {1\over 2}{\cal E}^{\mu \nu} {\cal E}^\tau R_{\mu\nu\tau\lambda}A^\lambda \cr
	= &-{1\over 2}
	(\epsilon^{\mu\nu\tau\rho}{\cal E}^{\pi}_{\rho}+ g^{\nu\tau} {\cal
		E}^\mu-g^{\mu\tau} {\cal E}^\nu) R_{\mu\nu\tau\lambda}A^\lambda \cr
	=&- g^{\nu\tau} {\cal E}^\mu R_{\mu\nu\tau\lambda}A^\lambda \cr =&-
	R_{\mu\lambda} {\cal E}^\mu A^\lambda , \cr}                       \eqno (9.12)
$$
which vanishes if $R_{\mu\lambda}= R_{\lambda\mu} = g^{\nu\tau}
R_{\tau\mu\nu\lambda}=0$. This result is independent of the tensor $A^\lambda$,
giving the gravitational field equations in vacuo
$$
{\cal D}_\wedge {\cal E}^\kappa =  {1\over 2}{\cal E}^{\mu \nu}(\partial_\mu \partial_\nu -
\partial_\nu\partial_\mu) {\cal E}^\kappa=0.                                  \eqno (9.13)
$$
This shows that the commutation of differentials corresponds to the vanishing of
the Ricci tensor, which is just Einstein's condition for the gravitational field equations.
In other words, {\it the algebraic formulation 
ensures that the components of the Riemann-Christoffel tensor satisfy
the field equations of general relativity}.

The square of algebraic invariant 
${\cal D}= \hat{\gamma}^{\pi 6}\hat{\gamma}^{\mu}\partial_\mu$ is
$$\eqalign{
{\cal D}^2 =& - {\cal E}^\mu \partial_\mu {\cal E}^\nu \partial_\nu 
= {\cal E}^\mu{\cal E}^\nu(\partial_\mu\partial_\nu+ \Gamma^\tau_{\mu \nu}\partial_\tau)\cr
=&\>(g^{\mu\nu} +{\cal E}^{\mu\nu}) 
(\partial_\mu\partial_\nu+\Gamma^\tau_{\mu \nu}\partial_\tau)\cr 
=& g^{\mu\nu}(\partial_\mu\partial_\nu+\Gamma^\tau_{\mu \nu}\partial_\tau) + {\cal D}_\wedge .}   \eqno (9.14)
$$
It was shown in \S3 that photon wave equations can be expressed in terms of a potential function 
${\bf A}$ that satisfies the Klein-Gordon equation corresponding
to the classical  equation relating the total energy $E=p_{*0}$ of a particle
to its mass and momentum, i.e $E^2 = {\vec p}^{\,2} + m^2 = p_\mu p^\mu$.
The Klein-Gordon equation in Riemannian space-time is obtained by replacing
$p_\mu\to \hat{\gamma}^{\pi 6}\partial_\mu$, and taking account of (9.13), to give
$$
{\cal D}^2 {\bf A}=g^{\mu\nu}(\partial_\mu\partial_\nu+\Gamma^\tau_{\mu \nu}\partial_\tau){\bf A} = 0.\eqno (9.15)
$$
This is the wave-equation for any zero rest mass boson. Photons only  
interact with charged particles and carry (algebraically) the information
required to make this distinction. Gravitons act on an any massive particle,
so that (9.15) provides their complete description, as far as can be achieved
in terms of the $Cl_{1,3}$ algebra.

It should be possible to express gravitational interactions in terms of the $Cl_{n,n}$
algebras by using the expression for unit time intervals obtained in this work,
but this has yet to investigated.

\beginsection \S10. Relationship with string theory 

String theories are based on adding additional spatial dimensions to
the three that are observed. This is often associated with extending 
the SO(1,3) algebra to SO$(1,q)$. Extending the Dirac algebra in the same
way provides a link with the Clifford algebras $Cl_{1,q}$. These are
only isomorphic with $Cl_{n,n}$ algebras if $q=1+8r$ where $r$ is a 
positive integer, e.g. $Cl_{1,9}\equiv Cl_{5,5}$ and  
$Cl_{1,17}\equiv Cl_{9,9}$. Only the physical interpretations of the 
$Cl_{5,5}$ sub-algebras of $Cl_{7,7}$ have been considered in this work. 

In order to relate CU with string theory 
the general notation for Clifford algebras, given in \S3, 
is compared with the labelling of $\gamma$ matrices 
used in Chapter 9 of [5]. That work denotes  the            
ten generators of $Cl_{1,9}$  $\gamma_i, i= 1,...,9,10$ 
where $\gamma_i^2 =-1$ for $i=1,2,...,9$, and $\gamma_{10}^2 =+1$.
The ten generators of $Cl_{5,5}$ will be labelled $\Gamma_i$, 
as in \S6, with the space-like generators
$\Gamma_i^2 =-1$ for $i=1,2,...,5$, and the time-like 
generators $\Gamma_i^2 =+1$ for $i= 6,...,10$.
The relationship between these generators follows 
that given on page 216 of [25], i.e.                                 
$$
\gamma^i ={\Gamma}_i{\bf h}, i=6,7,8,9 \>{\rm and} \>\gamma^i = \Gamma_i, i= 1,...,5,10 \eqno (10.1)
$$
where ${\bf h}= {\Gamma}_6{\Gamma}_7{\Gamma}_8{\Gamma}_9$. 
This makes it clear that the three space-like generators that
correspond to physical space are identical in algebraic and 
string theory. However, the single time-like generator 
$\gamma^{10}$ in $Cl_{1,9}$, 
associated with time in string theory, does not coincide 
with the time direction defined in this work.

In order to distinguish the five possible forms of 
ten-dimensional string theory,  
the number of dimensions have been extended to eleven by
including the  matrix $\gamma^{11}$ which, 
following equation (9.10) of [5], is defined as                        
$$
\gamma^{11}= \gamma^{10}\,\gamma^1\,\gamma^2...\gamma^9 = 
{\Gamma}_{10}{\Gamma}_1\,{\Gamma}_2\,{\Gamma}_3\,{\Gamma}_5\,
{\Gamma}_6{\bf h}\,{\Gamma}_7{\bf h}\,{\Gamma}_8{\bf h}\,{\Gamma}_9{\bf h} =
{\Gamma}_{10}{\Gamma}_1\,{\Gamma}_2\,{\Gamma}_3\,{\Gamma}_5
{\Gamma}_6{\Gamma}_7{\Gamma}_8{\Gamma}_9=\Gamma_0                             \eqno (10.2)
$$
This makes it apparent that $\gamma^{11}$ corresponds
to the time direction identified in this work, 
and which, as an operator, takes eigenvalues that 
distinguish between particles and anti-particles. 

\vskip100pt

\beginsection \S11. Substrates

It has been argued that physical substrates, described by the quantum numbers $\mu_B,\,\mu_C\,\mu_D,\,\mu_E,\,\mu_F,\,\mu_G$,
provide the medium for fermion wave-functions, and determine their properties.
Symmetry breaking determines possible fermion interactions, and correlates them 
with regions of space that have different substrates: \vskip3pt

\item {S1.} Fermions, with ${\mu_B=1}$, have equal and opposite charges, 
and time directions, to their corresponding anti-fermions, which have ${\mu_B=-1}$. 
Experimentally, anti-fermions are unstable in all accessible regions 
of space, suggesting that remote regions of space could exist in which 
anti-fermions are stable and fermions unstable.\vskip3pt

\item {S2.} The quantum number $\mu_C=i\mu_{\pi 6}= \pm 1$ distinguishes  
the two fermions in any doublet. The corresponding element of $Cl_{3,3}$ is 
$\hat\gamma^{\pi6}$, which is identified in \S3 and \S4 as providing the lepton substrate. 
Fermions or anti-fermions with ${\mu_C=-1}$ have one more charge than the ${\mu_C=+1}$ 
fermions or anti-fermions in the same doublet. $i\hat\gamma^{\pi6} $,  
$ i\hat\gamma^{\pi7}$ and $ i\hat\gamma^{\pi8}$ together generate the Lie algebra of SU(2),
defining an iso-spin algebra isomorphic to spin. If $i\hat\gamma^{\pi6} $ 
of this algebra is diagonal at all points in space-time,
this is analogous to the symmetry breaking in ferro-magnets, 
making the wave motion of leptons isomorphic with spin-waves. \vskip3pt

\item {S3.} The quantum numbers $\mu_D=\pm1,\,\mu_E=\pm1 $ together distinguish 
leptons and quarks, as shown in Table 7.1. They correspond to the commuting elements  
$\Gamma^D $ and $\Gamma^E$ of $Cl_{5,5}(LQ)$ and, combined with 
$\Gamma^X = -\Gamma^E \Gamma^D \Gamma^B $, determine the three commuting elements
of a sub-algebra, denoted $Cl_{3,3}(Q)$. Elements of this 
sub-algebra provide all 15 generators of the Lie algebra of SU(4) 
and its subgroup SU(3) that describes gluons. The SU(4)$\to$SU(3) symmetry breaking
is forced by the different charges on quarks and leptons
and distinguishes the substrate in `hadronic' space, produced by the gluons inside baryons 
and mesons, from the external `leptonic' space available only to leptons. \vskip3pt

\item {S4.} The quantum numbers $\mu_F=\pm1,\,\mu_G=\pm1 $ together distinguish four 
generations of leptons and quarks, as shown in Tables 8.1 and 8.2. 
They correspond to the commuting elements  
$\bar\Gamma^F $ and $\bar\Gamma^G$ of $Cl_{7,7}$ and, combined with 
$\bar\Gamma^H = -\bar\Gamma^F \bar\Gamma^G \bar\Gamma^C $, 
determine the three commuting elements of its sub-algebra, denoted $Cl_{3,3}(G)$.   
SU(4)$\to$SU(3) symmetry breaking is forced by the different charges
on fermions in the first to third generations  and on those in the fourth 
generation, as shown in Tables 8.1 and 8.2. In analogy with  the distinction 
between leptonic and hadronic regions of space described above, this suggests that
'dark matter' regions of space do not contain the substrate of 'ordinary matter', 
i.e. matter composed of fermions in the first three generations that are the 
constituents of solar systems.

\beginsection \S12. Conclusions

The starting point of this work was the integration of the 
macroscopic space-time algebra $Cl_{1,3}$, as developed in [7,8], with
the Dirac algebra, where it is treated as an invariant. It
was shown in \S3 that this is achieved with the $Cl_{3,3}$
algebra, producing the modified Dirac equation, which  
takes the form of a Lorentz invariant operator acting on
the 8-component Lorentz invariant column vector.  The physical
interpretation of lepton properties in  terms of $Cl_{3,3}$ in 
\S4 - \S6 then suggested extending the algebra to $Cl_{5,5}$ 
and $Cl_{7,7}$ in order to provide a description of all
known elementary fermions and their interactions.\vskip3pt 
 
Crucial features of the work are \vskip3pt
\item{1.} Identifying the proper time coordinate as the product of generators
for all the $Cl_{n,n},\>\{n=3,5,7\}$ algebras.\vskip3pt
\item{2.} Maintaining the algebraic distinction between
observers' space-time coordinate frames and fermion rest frames.\vskip3pt
\item{3.} Choosing the appropriate algebraic description for the physical 
space-time coordinates.\vskip3pt
\item{4.} Eliminating chiral symmetry breaking from the description weak 
interactions.\vskip3pt
\item{5.} Specifying all known elementary fermions in 
terms of seven binary quantum numbers. \vskip3pt 
\item{6.} Obtaining a formula for the charges on all known elementary fermions
in terms of the seven quantum numbers. \vskip3pt 
\item{7.} Relating the seven commuting elements of $Cl_{7,7}$ to 
different possible {\it substrates} for fermion and boson wave motion, and
showing that all elementary particle properties are determined by their substrate.\vskip3pt
\item{8.} Expressing the known gauge fields in terms of elements of $Cl_{7,7}$.\vskip3pt
\item{9.} Showing that the same {\it closure} property of $Cl_{1,3}$ 
determines the form of both the electromagnetic and gravitational 
field equations (\S3 and \S9).\vskip3pt 
\item{10.} Prediction of the existence of, and the charges on, a 4-th generation of fermions.\vskip5pt
This work remains incomplete, especially in relation to gravitation and the determination of fermion masses.
It does, nevertheless, provide a new starting point for further developments.

\beginsection  Acknowledgements

I am particularly grateful to Professor Ron King for his kind and patient help, 
given over many years, to correct my mathematical and conceptual errors. Thanks are also
due to Professor Ian Aitchison for his patience in pointing out errors in 
some of my assumptions many years ago.  I am also grateful to Professor Geoffrey Stedman 
for his continuing encouragement and support. 

\vfill\eject

\beginsection Appendix A: Representations of  $Cl_{3,3}$

The canonical $\gamma$-matrix representation of $Cl_{3,3}$ 
has 64 linearly independent real 8$\times$8 matrices. 
These representation matrices are expressed below as a 
multiplication table, which gives the products of
the representation matrices of the elements of $Cl_{1,3}$ (left
factors) with the unit matrix and matrices of the time-like generators
of $Cl_{3,3}$ (right factors). Each $\gamma$-matrix is expressed 
as a Kronecker product of three real $2\times 2$ matrices defined by
$$
{\bf I} =
\left(\matrix{1&0\cr
	0&1\cr}\right),\>
{\bf P} = -i\sigma_2 = \left(\matrix{0&-1\cr
	1&0\cr}\right),\>
{\bf Q} = \sigma_1 =\left(\matrix{0&1\cr
	1&0\cr}\right),\>
{\bf R} = -\sigma_3 =\left(\matrix{-1&0\cr
	0&1\cr}\right),     \eqno (A.1)                        
$$
where the $\sigma$s are the Pauli matrices. The real 
matrices satisfy the relations
$$
-{\bf P}^2 = {\bf Q}^2 = {\bf R}^2 = {\bf I}, \>\>
{\bf P}{\bf Q} = {\bf R} = -{\bf Q}{\bf P},\>{\bf P}{\bf R} = -{\bf
	Q} = -{\bf R}{\bf P}, \>{\bf Q}{\bf R} = -{\bf P} = -{\bf R}{\bf Q}.
\eqno (A.2)
$$
 
$$\vbox
{\settabs 5 \columns\+Table A1: Real "canonical" representation of $Cl_{3,3}$, which defines particle rest frames\cr
	\+||||||||||||||||||||||||||||||||||||||||||||\cr \+
	& $\>\>\>\>\>{\bf 1}\>$  &$\>\>\>\>\>\>\>\>\>\gamma^6$&
	$\>\>\>\>\>\>\>\>\>\gamma^7$&$\>\>\>\>\>\>\>\>\>\gamma^8 $\cr
	\+||||||||||||||||||||||||||||||||||||||||||||\cr
	\+$\>\>\>\>\>\>\>\bf 1$&$\>\>\>{\bf I}\otimes {\bf I}\otimes {\bf I}$&
	$\>\>\> {\bf I}\otimes {\bf Q }\otimes{\bf I}$&
	$-{\bf P}\otimes {\bf P}\otimes {\bf Q}$& $\>\>\>{\bf P}\otimes {\bf P}\otimes {\bf R}$\cr \+&&&\cr
	\+$\>\>\>\>\>\>\>\gamma^\pi $&$\>\> {\bf I}\otimes {\bf Q}\otimes{\bf P}$&
	$\>\> {\bf I}\otimes {\bf I}\otimes{\bf P}$&
	$\>\> {\bf P}\otimes {\bf R}\otimes{\bf R}$ &$\>\> {\bf P}\otimes {\bf R}\otimes{\bf Q}$\cr
	\+||||||||||||||||||||||||||||||||||||||||||||\cr
	\+$\>\>\>\>\>\>\>\gamma^0$&$-{\bf I}\otimes {\bf R}\otimes {\bf I}$&
	$ -{\bf I}\otimes {\bf P }\otimes{\bf I}$&
	$ \>\>\>{\bf P}\otimes {\bf Q}\otimes{\bf Q}$& $-{\bf P}\otimes {\bf Q}\otimes {\bf R}$\cr\+&&&\cr
	\+$\>\>\>\>\>\>\>\gamma^1$&$-{\bf Q}\otimes {\bf P}\otimes {\bf I}$&
	$- {\bf Q}\otimes {\bf R }\otimes{\bf I}$&
	$\>\>\>{\bf R}\otimes {\bf I}\otimes {\bf Q}$& $ -{\bf R}\otimes {\bf I}\otimes{\bf R}$\cr \+&&&\cr
	\+$\>\>\>\>\>\>\>\gamma^2$&$\>\>{\bf P}\otimes {\bf P}\otimes {\bf P }$&
	$\>\>{\bf P}\otimes {\bf R}\otimes {\bf P}$&
	$- {\bf I}\otimes {\bf I}\otimes{\bf R}$&$- {\bf I}\otimes {\bf I}\otimes{\bf Q}$ \cr\+&&&\cr
	\+$\>\>\>\>\>\>\>\gamma^3$&$\>\>\> {\bf R}\otimes {\bf P }\otimes{\bf I}$&
	$\>\> {\bf R}\otimes {\bf R }\otimes{\bf I}$&
	$\>\>\>{\bf Q}\otimes {\bf I}\otimes {\bf Q}$&$ -{\bf Q}\otimes {\bf I}\otimes{\bf R}$ \cr
	\+||||||||||||||||||||||||||||||||||||||||||||\cr
	\+$\>\>\>\>\>\>\>\gamma^{12}$&$-{\bf R}\otimes {\bf I}\otimes {\bf P}$&
	$- {\bf R}\otimes {\bf Q}\otimes{\bf P}$&
	$\>\> {\bf Q}\otimes {\bf P}\otimes{\bf R}$& $\>\>\>{\bf Q}\otimes {\bf P}\otimes {\bf Q}$\cr\+&&&\cr
	\+$\>\>\>\>\>\>\>\gamma^{31}$&$\>\> {\bf P}\otimes {\bf I}\otimes{\bf I}$&
	$\>\> {\bf P}\otimes {\bf Q}\otimes{\bf I}$&
	$\>\> {\bf I}\otimes {\bf P}\otimes{\bf Q}$ &$- {\bf I}\otimes {\bf P}\otimes{\bf R}$\cr\+&&&\cr
	\+$\>\>\>\>\>\>\>\gamma^{23}$&$\>\> {\bf Q}\otimes {\bf I}\otimes{\bf P}$&
	$\>\> {\bf Q}\otimes {\bf Q}\otimes{\bf P}$&
	$\>\>{\bf R}\otimes {\bf P}\otimes {\bf R}$ &$\>\> {\bf R}\otimes {\bf P}\otimes{\bf Q}$\cr\+&&&\cr
	\+$\>\>\>\>\>\>\>\gamma^{03}$&$- {\bf R} \otimes {\bf Q}\otimes{\bf I}$&
	$- {\bf R} \otimes {\bf I}\otimes{\bf I}$&
	$- {\bf Q}\otimes {\bf R}\otimes{\bf Q}$ &$\>\> {\bf Q}\otimes {\bf R}\otimes{\bf R}$\cr\+&&&\cr
	\+$\>\>\>\>\>\>\>\gamma^{02}$&$- {\bf P}\otimes {\bf Q}\otimes{\bf P}$&
	$- {\bf P}\otimes {\bf I}\otimes{\bf P}$&
	$\>\>\> {\bf I}\otimes {\bf R}\otimes{\bf R}$ &$\>\> {\bf I}\otimes {\bf R}\otimes{\bf Q}$\cr\+&&&\cr
	\+$\>\>\>\>\>\>\>\gamma^{01}$&$\>\>{\bf Q}\otimes {\bf Q}\otimes {\bf I}$&
	$\>\>{\bf Q}\otimes {\bf I}\otimes {\bf I}$&
	$- {\bf R}\otimes {\bf R}\otimes{\bf Q}$ &$\>\>{\bf R}\otimes {\bf R}\otimes {\bf R}$\cr
	\+||||||||||||||||||||||||||||||||||||||||||||\cr
	\+$\>\>\>\>\>\>\>\gamma^{\pi 0}$&$\>\>\>{\bf I}\otimes {\bf P}\otimes {\bf P}$&
	$\>\>\>{\bf I}\otimes {\bf R}\otimes {\bf P}$&
	$\>\> {\bf P}\otimes {\bf I}\otimes{\bf R}$ &$\>\>\> {\bf P}\otimes {\bf I}\otimes{\bf Q}$\cr\+&&&\cr
	\+$\>\>\>\>\>\>\>\gamma^{\pi 1}$&$\>\> {\bf Q}\otimes {\bf R}\otimes{\bf P}$&
	$\>\>{\bf Q}\otimes {\bf P}\otimes{\bf P}$&
	$\>\> {\bf R}\otimes {\bf Q}\otimes{\bf R}$ &$\>\> {\bf R}\otimes {\bf Q}\otimes{\bf Q}$\cr\+&&&\cr
	\+$\>\>\>\>\>\>\>\gamma^{\pi 2}$&$\>\> {\bf P}\otimes {\bf R}\otimes{\bf I}$&
	$\>\>\> {\bf P}\otimes {\bf P}\otimes{\bf I}$&
	$\>\>\>\>{\bf I}\otimes {\bf Q}\otimes{\bf Q}$&$ -{\bf I}\otimes {\bf Q}\otimes{\bf R}$ \cr\+&&&\cr
	\+$\>\>\>\>\>\>\>\gamma^{\pi 3}$&$- {\bf R}\otimes {\bf R}\otimes{\bf P}$&
	$-{\bf R}\otimes {\bf P}\otimes {\bf P}$&
	$\>\> {\bf Q}\otimes {\bf Q}\otimes{\bf R}$ &$\>\>{\bf Q}\otimes {\bf Q}\otimes {\bf Q}$\cr
	\+||||||||||||||||||||||||||||||||||||||||||||\cr
	\+||||||||||||||||||||||||||||||||||||||||||||\cr}
$$
  
The 64 $\hat\gamma$-matrix representation of $Cl_{3,3}$ given in Table A2 is obtained using 
a transformation of the canonical representation matrices that makes both $\gamma^{56}$ and $
\gamma^{12}$ diagonal. Defining
${\bf Z} = {1\over \sqrt 2}(-{\bf R} + i {\bf P}) $ gives
$$
{\bf Z}{\bf P}{\bf Z}^{-1}= i{\bf R},\> {\bf Z}{\bf Q}{\bf Z}^{-1}=
-{\bf Q},\>{\bf Z}{\bf R}{\bf Z}^{-1}= -i{\bf P},\>\>{\bf Z}^2 =
{\bf I},\> {\bf Z}^{-1}={\bf Z}^\dagger = {\bf Z}. \eqno (A.3)
$$
It follows that the transformation $\hat \gamma ={\cal Z}\gamma{\cal Z}^{-1}$,
where ${\cal Z} = {\bf Z}\otimes {\bf I}\otimes{\bf Z} $, transforms 
real matrices in the canonical representation in Table A1 to the complex matrices of the
modified canonical representation $\hat\gamma$ given below. 
$$\vbox
{\settabs 5 \columns\+ Table A2: The $\hat{\gamma}$ fermion rest frame representation 
	of $Cl_{3,3}$ \cr
	 \+||||||||||||||||||||||||||||||||||||||||||||\cr \+ &
	$\>\>\>\>\>{\bf 1}\> $  &$\>\>\>\>\>\>\>\>\>\hat\gamma^6$&
	$\>\>\>\>\>\>\>\>\>\hat\gamma^7$&$\>\>\>\>\>\>\>\>\>\hat\gamma^8 $\cr
	\+||||||||||||||||||||||||||||||||||||||||||||\cr \+$\>\>\>\>\>\>\>{\bf
	1}_3$&$\>\>\>{\bf I}\otimes {\bf I}\otimes {\bf I}$& $\>\>\> {\bf I}\otimes {\bf Q
	}\otimes{\bf I}$& $\>\>\>i{\bf R}\otimes {\bf P}\otimes {\bf Q}$& $\>\>{\bf
	R}\otimes {\bf P}\otimes {\bf P}$\cr \+&&&\cr
\+$\>\>\>\>\>\>\>\hat\gamma^\pi $&$\>\> i{\bf I}\otimes {\bf Q}\otimes{\bf R}$&
$\>\> i{\bf I}\otimes {\bf I}\otimes{\bf R}$& $\> {\bf R}\otimes {\bf
	R}\otimes{\bf P}$ &$\> -i{\bf R}\otimes {\bf R}\otimes{\bf Q}$\cr
\+||||||||||||||||||||||||||||||||||||||||||||\cr \+$\>\>\>\>\>\>\>\hat\gamma^0
$&$-{\bf I}\otimes {\bf R}\otimes {\bf I}$& $ -{\bf I}\otimes {\bf P
}\otimes{\bf I}$& $-i{\bf R}\otimes {\bf Q}\otimes{\bf Q}$& $-{\bf R}\otimes
{\bf Q}\otimes {\bf P}$\cr\+&&&\cr \+$\>\>\>\>\>\>\>\hat\gamma^1$&${\bf
	Q}\otimes {\bf P}\otimes {\bf I}$& ${\bf Q}\otimes {\bf R }\otimes{\bf I}$&
$-i{\bf P}\otimes {\bf I}\otimes {\bf Q}$& $ {\bf P}\otimes {\bf I}\otimes{\bf
	P}$\cr \+&&&\cr \+$\>\>\>\>\>\>\>\hat\gamma^2$&$\>\>-{\bf R}\otimes {\bf
	P}\otimes {\bf R}$& $\>\>-{\bf R}\otimes {\bf R}\otimes {\bf R}$& $\>\>\>i{\bf
	I}\otimes {\bf I}\otimes{\bf P}$&$ {\bf I}\otimes {\bf I}\otimes{\bf Q}$
\cr\+&&&\cr \+$\>\>\>\>\>\>\>\hat\gamma^3 $&$\>\>\> -i{\bf P}\otimes {\bf P
}\otimes{\bf I}$& $\>\>-i {\bf P}\otimes {\bf R }\otimes{\bf I}$& ${\bf
Q}\otimes {\bf I}\otimes {\bf Q}$&$ -i{\bf Q}\otimes {\bf I}\otimes{\bf P}$ \cr
\+||||||||||||||||||||||||||||||||||||||||||||\cr
\+$\>\>\>\>\>\>\>\hat\gamma^{12}$&$-{\bf P}\otimes {\bf I}\otimes {\bf R}$&
$-{\bf P}\otimes {\bf Q}\otimes{\bf R}$& $\>i {\bf Q}\otimes {\bf P}\otimes{\bf
	P}$& $\>\>{\bf Q}\otimes {\bf P}\otimes {\bf Q}$\cr\+&&&\cr
\+$\>\>\>\>\>\>\>\hat\gamma^{31}$&$\>\> i{\bf R}\otimes {\bf I}\otimes{\bf I}$&
$\>\> i{\bf R}\otimes {\bf Q}\otimes{\bf I}$& $\>- {\bf I}\otimes {\bf
	P}\otimes{\bf Q}$ &$i {\bf I}\otimes {\bf P}\otimes{\bf P}$\cr\+&&&\cr
\+$\>\>\>\>\>\>\>\hat\gamma^{23}$&$\>\> -i{\bf Q}\otimes {\bf I}\otimes{\bf R}$&
$\>\> -i{\bf Q}\otimes {\bf Q}\otimes{\bf R}$& $\>-{\bf P}\otimes {\bf P}\otimes
{\bf P}$ &$\> i{\bf P}\otimes {\bf P}\otimes{\bf Q}$\cr\+&&&\cr
\+$\>\>\>\>\>\>\>\hat\gamma^{03}$&$ i{\bf P} \otimes {\bf Q}\otimes{\bf I}$&
$i{\bf P} \otimes {\bf I}\otimes{\bf I}$& $\>-{\bf Q}\otimes {\bf R}\otimes{\bf
	Q}$ &$\>i {\bf Q}\otimes {\bf R}\otimes{\bf P}$\cr\+&&&\cr
\+$\>\>\>\>\>\>\>\hat\gamma^{02}$&${\bf R}\otimes {\bf Q}\otimes{\bf R}$& $ {\bf
	R}\otimes {\bf I}\otimes{\bf R}$& $\>\>-i {\bf I}\otimes {\bf R}\otimes{\bf P}$
&$\>- {\bf I}\otimes {\bf R}\otimes{\bf Q}$\cr\+&&&\cr
\+$\>\>\>\>\>\>\>\tilde\gamma^{01}$&$-{\bf Q}\otimes {\bf Q}\otimes {\bf I}$&
$-{\bf Q}\otimes {\bf I}\otimes {\bf I}$& $-i {\bf P}\otimes {\bf R}\otimes{\bf
	Q}$ &$\>-{\bf P}\otimes {\bf R}\otimes {\bf P}$\cr
\+||||||||||||||||||||||||||||||||||||||||||||\cr
\+$\>\>\>\>\>\>\>\hat\gamma^{\pi 0}$&$\>\>\>i{\bf I}\otimes {\bf P}\otimes {\bf
	R}$& $\>\>i{\bf I}\otimes {\bf R}\otimes {\bf R}$& $\>\> {\bf R}\otimes {\bf
	I}\otimes{\bf P}$ &$\>\>-i {\bf R}\otimes {\bf I}\otimes{\bf Q}$\cr\+&&&\cr
\+$\>\>\>\>\>\>\>\hat\gamma^{\pi 1}$&$-i{\bf Q}\otimes {\bf R}\otimes{\bf R}$&
$\>-i{\bf Q}\otimes {\bf P}\otimes{\bf R}$& $\>- {\bf P}\otimes {\bf
	Q}\otimes{\bf P}$ &$\>-i {\bf P}\otimes {\bf Q}\otimes{\bf Q}$\cr\+&&&\cr
\+$\>\>\>\>\>\>\>\hat\gamma^{\pi 2}$&$\>\> i{\bf R}\otimes {\bf R}\otimes{\bf I}$&
$\>\>\> i{\bf R}\otimes {\bf P}\otimes{\bf I}$& $\>\>\>-{\bf I}\otimes {\bf
	Q}\otimes{\bf Q}$&$ i{\bf I}\otimes {\bf Q}\otimes{\bf P}$ \cr\+&&&\cr
\+$\>\>\>\>\>\>\>\hat\gamma^{\pi 3}$&$- {\bf P}\otimes {\bf R}\otimes{\bf R}$&
$-{\bf P}\otimes {\bf P}\otimes {\bf R}$& $\>\>i {\bf Q}\otimes {\bf
	Q}\otimes{\bf P}$ &$\>\>{\bf Q}\otimes {\bf Q}\otimes {\bf Q}$\cr
\+||||||||||||||||||||||||||||||||||||||||||||\cr
\+||||||||||||||||||||||||||||||||||||||||||||\cr} $$

The  matrix  representations in Tables A1 and A2 relate to fermion rest frames.
Representation matrices
for arbitrary reference frames  are obtained by
 Lorentz transformations 
$\gamma \rightarrow {\bf\Lambda}\gamma{\bf\Lambda}^{-1.}$, 
where $\bf\Lambda$ is defined in (3.13).
Relationships between the various 4$\times$4
matrix representations of fermion rest frame coordinate systems
are given in Table A3.

$$\vbox
{\settabs 6\columns\+Table A3: Alternative  choices of space-time representation matrices  \cr
	\+||||||||||||||||||||||||||||||||||||||||||||\cr \+
	& $\>\>\>\>\>\>\gamma\>$  &$\>\>\>\>\>\>\bar\gamma$&$\>\>\>\>\>\>\hat{\gamma}$&
	$\>\>\>\>\>\>{^a\gamma}$&$\>\>\>\>\>\>{^b\gamma} $&\cr
	\+||||||||||||||||||||||||||||||||||||||||||||\cr
	\+$\>\>\>\>\>\>\gamma^0$&$-{\bf I}\otimes {\bf R}\otimes {\bf I}$&$\> -{\bf I}\otimes {\bf R }$&
	$ \>-{\bf I}\otimes {\bf R}\otimes {\bf I}$& $\>-{\bf I}\otimes {\bf R}$&$-{\bf I}\otimes{\bf R}$\cr
	\+$\>\>\>\>\>\>\gamma^1$&$-{\bf Q}\otimes {\bf P}\otimes {\bf I}$&$- {\bf Q}\otimes {\bf P }$&
	$\>\>\>{\bf Q}\otimes {\bf P}\otimes {\bf I}$& $\>\>\> {\bf Q}\otimes {\bf P}$&$\>\>\>{\bf Q}\otimes {\bf P}$\cr 
	\+$\>\>\>\>\>\>\gamma^2$&$\>\>\>{\bf P}\otimes {\bf P}\otimes {\bf P }$&$-i{\bf P}\otimes {\bf P}$&
	$- {\bf R}\otimes {\bf P}\otimes{\bf R}$&$\>\>\>{\bf R}\otimes {\bf P}$&$-{\bf R}\otimes {\bf P}$\cr
	\+$\>\>\>\>\>\>\gamma^3$&$\>\>\> {\bf R}\otimes {\bf P }\otimes{\bf I}$&$\>\> \>\>\>{\bf R}\otimes {\bf P }$&
	$-i{\bf P}\otimes {\bf P}\otimes {\bf I}$&$ -i{\bf P}\otimes {\bf P}$ &$-i{\bf P}\otimes {\bf P}$\cr
	\+||||||||||||||||||||||||||||||||||||||||||||\cr
	\+$\>\>\>\>\>\>\gamma^\pi $&$\>\>\>\> {\bf I}\otimes {\bf Q}\otimes{\bf P}$&$-i {\bf I}\otimes {\bf Q}$&
	$\>\> i{\bf I}\otimes {\bf Q}\otimes{\bf R}$ &$\>-i {\bf I}\otimes {\bf Q}$&$\>\>\>\>\>i{\bf I}\otimes{\bf Q}$\cr
	\+||||||||||||||||||||||||||||||||||||||||||||\cr}
$$

\beginsection Appendix B. Block diagonalized representations
 
The modified canonical representations $\hat{\gamma}$ 
puts structors into block diagonal form.  
The $\hat{\gamma}$ representation of the differential 
structor $\bf D$ is
$$
{\bf D}  = \hat\gamma^{\mu} \partial_{\mu}
= \left(\matrix{ {\bf D}_a&0\cr
	0& {\bf D}_b\cr}
\right) \eqno (B.1)
$$
where 
$$
{\bf D}_a  =
\left(\matrix{\partial_0&0&\partial_2&-\partial_1-i\partial_3\cr
	0&\partial_0&-\partial_1+i\partial_3&-\partial_2\cr
	-\partial_2&\partial_1+i\partial_3&-\partial_0&0\cr
	\partial_1-i\partial_3&\partial_2&0&-\partial_0\cr}
\right) \eqno (B.1a)
$$
and
$$
{\bf D}_b  =
\left(\matrix{\partial_0&0&-\partial_2&-\partial_1-i\partial_3\cr
	0&\partial_0&-\partial_1+i\partial_3&\partial_2\cr
	\partial_2&\partial_1+i\partial_3&-\partial_0&0\cr
	\partial_1-i\partial_3&-\partial_2&0&-\partial_0\cr}
\right)\eqno (B.1b)
$$

The  general potential structor has the $\hat{\gamma}$ 
block diagonal representation
$$
{\bf A}  = \hat\gamma^{\mu} (A_{\mu} - \hat\gamma^\pi A_{\pi\mu}) 
= \left(\matrix{ {\bf A}_a&0\cr
	0& {\bf A}_b\cr}
\right),                                                       \eqno (B.2)
$$
where 
$$
{\bf A}_a  =
\left(\matrix{A_0+iA_{\pi2}&A_{\pi3}-iA_{\pi1}&A_2+iA_{\pi0}&-A_1-iA_3\cr
	-A_{\pi3}-iA_{\pi1}&A_0-iA_{\pi2}&-A_1+iA_3&-A_2+iA_{\pi0}\cr
	-A_2-iA_{\pi0}&A_1+iA_3&-A_0-iA_{\pi2}&-A_{\pi3}+iA_{\pi1}\cr
	A_1-iA_3&A_2-iA_{\pi0}&A_{\pi3}+iA_{\pi1}&-A_0+iA_{\pi2}\cr}
\right)                                                        \eqno(B.2a)
$$
and
$$
{\bf A}_b  =
\left(\matrix{A_0+iA_{\pi2}&-A_{\pi3}-iA_{\pi1}&-A_2+iA_{\pi0}&-A_1-iA_3\cr
	A_{\pi3}+iA_{\pi1}&A_0-iA_{\pi2}&-A_1+iA_3&A_2-iA_{\pi0}\cr
	A_2+iA_{\pi0}&A_1+iA_3&-A_0-iA_{\pi2}&A_{\pi3}-iA_{\pi1}\cr
	A_1-iA_3&-A_2+iA_{\pi0}&-A_{\pi3}-iA_{\pi1}&-A_0+iA_{\pi2}\cr}
\right).                                                        \eqno(B.2b)
$$

Similarly, the field structor has 
the block diagonal $\hat{\gamma}$ matrix representation
$$
{\bf F}  = \hat\gamma^{\mu\nu} F_{\mu\nu} 
= \left(\matrix{ {\bf F}_a&0\cr
	0& {\bf F}_b\cr}
\right),                                                             \eqno (B.3)
$$
where
$$
{\bf F}_a = 
\left(\matrix{-iF_{31}&-F_{12}+iF_{23}& F_{02}        & F_{01}-iF_{03}\cr
	F_{12}+iF_{23} &iF_{31}           & F_{01}+iF_{03}&-F_{02}\cr
           F_{02}  &F_{01}-iF_{03}    & -iF_{31}      &-F_{12}+iF_{23}\cr
	F_{01}+iF_{03} &-F_{02}           & F_{12}+iF_{23}& iF_{31} \cr}\right)  \eqno (B.3a)        
$$
and
$$
{\bf F}_b =
\left(\matrix{-iF_{31}&F_{12}-iF_{23}& -F_{02}       & F_{01}-iF_{03}\cr
       -F_{12}-iF_{23}&iF_{31}       & F_{01}+iF_{03}&F_{02}\cr
	-F_{02}           &F_{01}-iF_{03}& -iF_{31}      &F_{12}-iF_{23}\cr
	F_{01}+iF_{03}    &F_{02}        &-F_{12}-iF_{23}& iF_{31} \cr}\right).  \eqno (B.3b)      
$$
As Lorentz transformations are also  expressed in terms of the 
matrices $\hat{\gamma}^{\mu\nu}$, they also have block diagonal form, viz.
$$
\Lambda   
= \left(\matrix{ \Lambda_a&0\cr
	0& \Lambda_b\cr}
\right).                                                             \eqno (B.4)
$$
\vskip 20pt

\beginsection References 

\frenchspacing
\item {[1]} Bettini, Alessandro 2008 Introduction to Elementary
Particle Physics (Cambridge University Press) 

\item {[2]}  Georgi, Howard 1982 Lie Algebras in Particle Physics (Benjamin
Publishing Co. Inc, Massachusetts)    

\item {[3]} Baez, John and Huerta, John 2010 The Algebra of Grand 
Unified Theories {arXiv:0904.1556v2}  

\item{[4]} Aitchison, Ian J. R. 2007 Supersymmetry in Particle Physics (Cambridge University Press)

\item {[5]} Schomerus, Volker  2017 A Primer on String Theory
(Cambridge University Press)  

\item {[6]} Eddington, Sir A.S. 1946 Fundamental Theory (Cambridge University Press)

\item {[7]} Hestenes, David 1966 Space-time algebra (Gordon and Breach, New York)  

\item {[8]} Doran, Chris and Lasenby, Anthony  2003 Geometric Algebra for Physicists
(Cambridge University Press) 

\item {[9]} Newman, D. J. 1958 Structure Theory
{\it Proc. Roy. Irish Acad.} {\bf 59}, 29-47     

\item {[10]} Trayling, Greg and Baylis, W. E. 2001  
A geometric basis for the standard-model gauge group 
{\it J. Phys. A: Math. Gen.} {\bf 34}, 3309-3324
   
\item {[11]} Dartora, C. A. and Cabrera, G. G. 2009 The Dirac equation and a non-chiral electroweak theory
in six dimensional space-time from a locally gauged $SO(3,3)$ symmetry group {arXiv: 0901.4230v1} 
Int J Theor Phys (2010) {\bf 49}:51-61                                                             

\item {[12]} \.Zenczykowski, P. 2009  Clifford algebra of
non-relativistic phase space and the concept of mass 
{\it J.Phys.A: Math.Theor.} {\bf 42}, 045204                 

\item {[13]} \.Zenczykowski, P. 2015  From Clifford algebra of
Nonrelativistic Phase Space to Quarks and Leptons of the Standard Model 
{\it Adv. Appl. Clifford Algebras} Springerlink.com 2015 DOI 10.1007/s00006-015-0564-7

\item {[14]} \.Zenczykowski, P. 2018 Quarks, Hadrons and Emergent Spacetime 
              {arXiv:1809.05402v1} 

\item {[15]} Stoica, O. C. 2018  The Standard Model Algebra:
Leptons, Quarks and Gauge from the Complex 
Clifford Algebra Cl$_6$ {\it Adv. Appl. Clifford Algebra}
{\bf 28}, 52. {arXiv:1702.04336v3} 

\item {[16]} Stoica, O. C. 2020  Chiral asymmetry in the weak interaction 
via Clifford Algebras {arXiv:2005.08855v1}  

\item {[17]} Pav\v si\v c, Matej 2021 Clifford Algebras, Spinors and Cl(8,8) Unification {arXiv:2105.11808}

\item {[18]} Yamatsu, Naoki 2020 USp(32) Special Grand Unification {arXiv:2007.08067v1}  

\item {[19]} Aitchison, I. J. R. and Hey, A. J. G. 2003  Gauge Theories in
Particle Physics, Volume I: From Relativistic Quantum Mechanics to QED (Taylor and Francis)

\item {[20]} Aitchison, I. J. R. and Hey, A. J. G. 2004  Gauge Theories in
Particle Physics, Volume II: QCD and the Electroweak Theory (IOP Publishing Ltd)    

\item {[21]} Thomson, Mark. 2013 Modern Particle Physics (Cambridge University Press)                                      

\item {[22]} Dodd, James and Gripalos, Ben 2020 The Ideas of Particle Physics
(Cambridge University Press)

\item {[23]} Hill, E. L. and Landshoff, R. 1938 The Dirac Electron Theory 
{\it Rev. Mod. Physics} {\bf 10}, 87-132                                   

\item {[24]} Newman, D. J. and Kilmister, C. W. 1959 A New Expression for 
Einstein's Law of Gravitation {\it Proc. Camb. Phil. Soc.} {\bf 55}, 139-141                                                                     

\item {[25]} Lounesto, Pertti 1997 Clifford Algebras and Spinors (Cambridge University Press) 

\end